\documentclass[fleqn,usenatbib,referee]{mnras}

\usepackage{newtxtext,newtxmath}

\usepackage[T1]{fontenc}

\DeclareRobustCommand{\VAN}[3]{#2}
\let\VANthebibliography\thebibliography
\def\thebibliography{\DeclareRobustCommand{\VAN}[3]{##3}\VANthebibliography}

\usepackage{graphicx}	
\usepackage{amsmath}	
\usepackage{threeparttable} 

\title[Probable Dormant neutron star in a short-period binary]{Probable Dormant Neutron Star in a Short-Period Binary System}

\author[T. Mazeh et al.]{
Tsevi Mazeh,$^{1}$\thanks{E-mail: mazeh@tauex.tau.ac.il}
Simchon Faigler,$^{1}$ 
Dolev Bashi,$^{1}$ 
Sahar Shahaf,$^{1,2}$
Niv Davidson,$^{1}$
Matthew Green,$^{1}$
\newauthor
Roy Gomel,$^{1}$
Dan Maoz,$^{1}$
Amitay Sussholz,$^{1}$
Subo Dong,$^{3}$
Haotong Zhang,$^{4}$
Jifeng Liu,$^{4}$
Song Wang,$^{4}$
\newauthor
Ali Luo,$^{4}$  
Zheng Zheng,$^{5}$ 
Na'ama Hallakoun,$^{2}$
Volker Perdelwitz,$^{6,7}$
David W. Latham,$^{8}$ 
Ignasi Ribas,$^{9,10}$ 
\newauthor
David Baroch,$^{9,10}$
Juan Carlos Morales,$^{9,10}$
Evangelos Nagel,$^{7,11}$
Nuno C. Santos,$^{12,13}$
David R. Ciardi,$^{14}$
\newauthor
Jessie L. Christiansen,$^{14}$
Michael B. Lund,$^{14}$ 
Joshua N.\ Winn$^{15}$ \\
$^{1}$School of Physics and Astronomy, Tel Aviv University, Tel Aviv, 6997801, Israel\\
$^{2}$Department of particle physics and astrophysics, Weizmann Institute of Science, Rehovot 7610001, Israel\\
$^{3}$Kavli Institute for Astronomy and Astrophysics, Peking University, Yi He Yuan Road 5, Hai Dian District, Beijing 100871, People's Republic of China\\
$^{4}$National Astronomical Observatories, CAS, People's Republic of China\\
$^{5}$Department of Physics and Astronomy, University of Utah, 115 South 1400 East, Salt Lake City, UT 84112, USA\\
$^{6}$Department of physics, Ariel University\\
$^{7}$Hamburger Sternwarte, Universität Hamburg, Gojenbergsweg 112, 21029 Hamburg, Germany\\
$^{8}$Center for Astrophysics Harvard \& Smithsonian, 60 Garden St., Cambridge, Mass., USA\\
$^{9}$Institut de Ciències de l'Espai (ICE, CSIC), Campus UAB, c/ Can Magrans s/n, E-08193 Bellaterra, Barcelona, Spain\\
$^{10}$Institut d'Estudis Espacials de Catalunya (IEEC), c/ Gran Capità 2- 4, E-08034 Barcelona, Spain\\
$^{11}$Thüringer Landessternwarte Tautenburg, Sternwarte 5, 07778 Tautenburg, Germany\\
{$^{12}$ Instituto de Astrof\'isica e Ci\^encias do Espa\c{c}o, Universidade do Porto, CAUP, Rua das Estrelas, 4150-762 Porto, Portugal}\\
{$^{13}$Departamento de F\'isica e Astronomia, Faculdade de Ci\^encias, Universidade do Porto, Rua do Campo Alegre, 4169-007 Porto, Portugal}\\
$^{14}$NASA Exoplanet Science Institute –-- Caltech IPAC\\
$^{15}$Department of Astrophysical Sciences, Princeton University, Princeton, NJ 08540, USA
}

\date{Accepted XXX. Received YYY; in original form ZZZ}

\pubyear{2015}

\begin{document}
\label{firstpage}
\pagerange{\pageref{firstpage}--\pageref{lastpage}}
\maketitle

\begin{abstract}
We have identified 2XMM J125556.57+565846.4, at a distance of 
600 pc, as a binary system consisting of a normal star and a probable dormant neutron star.  Optical spectra exhibit a slightly evolved F-type single star, displaying periodic Doppler shifts with a 2.76-day Keplerian circular orbit, with no indication of light from a secondary component.  Optical and UV photometry reveal ellipsoidal
modulation
with half the orbital period, due to the tidal deformation of the F star. 
The mass of the unseen companion is constrained to the range of
$1.1$--$2.1\, M_{\odot}$
at $3\sigma$ confidence, with the median of the mass distribution at $1.4\, M_{\odot}$, the typical mass of known neutron stars. A main-sequence star cannot masquerade as the dark companion. 
The distribution of possible companion masses still allows for the possibility of a very massive white dwarf. The companion itself could also be a close pair consisting of a white dwarf and an M star, or two white dwarfs, although the binary evolution that would lead to such a close triple system is unlikely.
Similar ambiguities regarding the certain identification of a dormant neutron star are bound to affect most future discoveries of this type of non-interacting system. 
If the system indeed contains a dormant neutron star, it will become, in the future, a bright X-ray source and 
afterwards
might even host a millisecond pulsar.
\end{abstract}



\section{Introduction}

Neutron stars are among the most exotic known astronomical objects, with typical masses of $1.2$--$2 \, M_{\odot}$ \citep[e.g.,][]{antoniadis16,ozel16} and radii of $\sim10$ km, resulting in mass densities on the order of $10^{14}$ g/cc \citep[e.g.,][]{ozel16}, similar to atomic-nucleus density.
They are formed by 
supernova explosions of stars with initial masses larger than $\sim8\, M_{\odot}$, and therefore are quite rare \citep{tauris06}.

Most of the known neutron stars are young pulsars, discovered by virtue of their periodic radio pulses, which reveal their fast rotation \citep[e.g.,][]{kaspi06}. Neutron stars cease emitting radio pulses after much of their rotational spin energy has been radiated away \citep{bransgrove18}. 
However, a neutron star can be revived as a pulsar if it resides in a binary system. When the companion leaves the main sequence (MS), its outer layers expand until they are pulled onto the neutron star, if the two stars are close enough.  The mass transferred onto the compact object can generate a high X-ray luminosity, while the associated angular momentum increases the spin rate of the neutron star \citep{vanden91}. Indeed, some neutron stars are visible as bright X-ray sources \citep[e.g.,][]{paul17}, and many others as 'revived' millisecond pulsars \citep{manchester17}.

During the long MS lifetime of the companion, the neutron star is extremely faint in all electromagnetic bands, making this type of system difficult to discover. The binary appears as a single optical-band star, without any X-ray or radio emission.
The binarity of such a system can only be revealed through gravitational interactions
%
%
with the unseen, relatively massive, companion.
%
In such cases, however, one must rule out the possibility that the companion is a fainter MS star, before concluding that the companion is a compact stellar remnant: a white dwarf, a neutron star, or a black hole. Most if not all recent efforts to detect compact binary companions in this manner \citep{LB1_20,HR6819_20}
have not been able to definitively rule out the possibility that the secondary is a MS or a sub-giant star, \citep[e.g.,][]{mazeh20,el-badry22a,el-badry22b}.

In this paper, we identify 2XMM J125556.57+565846.4, hereafter J1255, a slightly evolved F-type star,
as a likely MS plus neutron star system. The system was selected based on large radial-velocity (RV) 
differences 
(${\rm RV}_{\rm max}-{\rm RV}_{\rm min}$)
seen in the LAMOST stellar RV public database. Follow-up spectroscopic observations reveal 
J1255 as a 2.76-day single-lined spectroscopic binary system.
{\it TESS} optical and Swift-UVOT ultraviolet photometry display  ellipsoidal %
modulation
with half the orbital period, due to the tidal deformation of the F star. 
The data constrain the mass of the unseen companion to a range 
$1.1$--$2.1\, M_{\odot}$  ($3\sigma$), consistent with the typical mass of the known neutron stars.
This is the first dormant neutron-star candidate discovered in a short-period binary.

Just before submission of the paper, the {\it Gaia} large sample of spectroscopic binaries became available \citep{NSS}. The catalog includes an orbit for J1255,  Gaia DR3 1577114915964797184, consistent with our spectroscopic orbit. 

Section~\ref{sec:RV} presents the radial-velocity observations of the system, Section~\ref{sec:photometry} the optical and UV photometry of the binary, Section~\ref{sec:stellar_parameters} derives the stellar parameters of the primary and Section~\ref{sec:Mass} derives the companion mass. We discuss our results in Section~\ref{sec:discussion}.

\section{Radial velocities of J1255}
\label{sec:RV}

\subsection{LAMOST radial velocities}

\begin{figure}
\centering
\includegraphics[height=7cm,width=8cm]{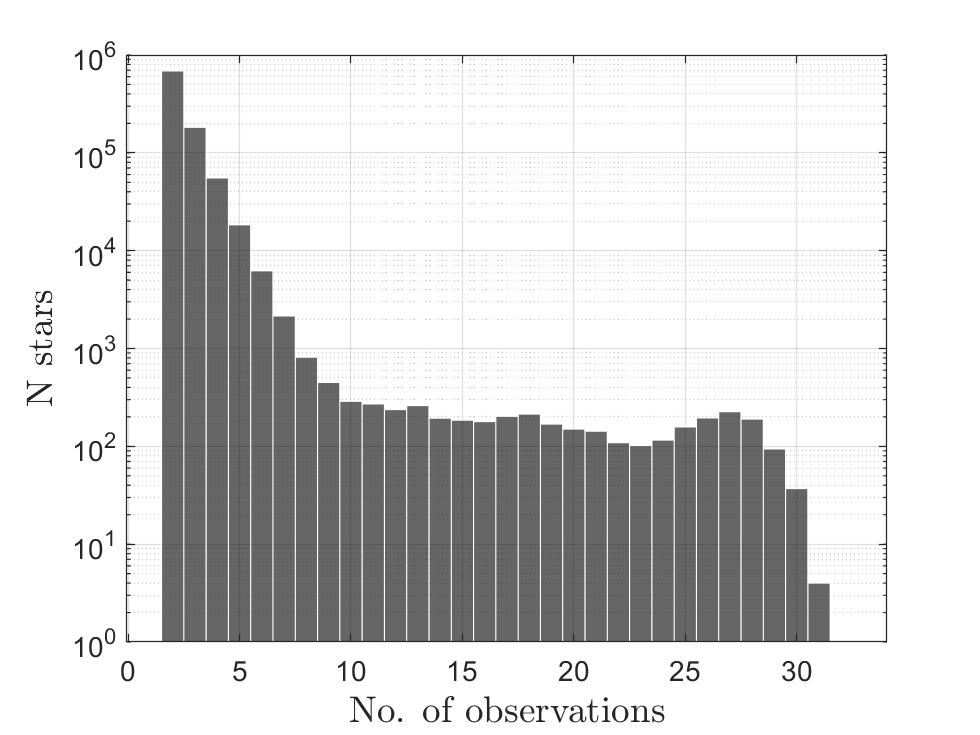}
\caption{Distribution of number of RV epochs for stars with multiple measurements in LAMOST DR6}
\label{fig:histogram}
\end{figure}

\begin{figure}
\centering
\includegraphics[height=7cm,width=8cm]
{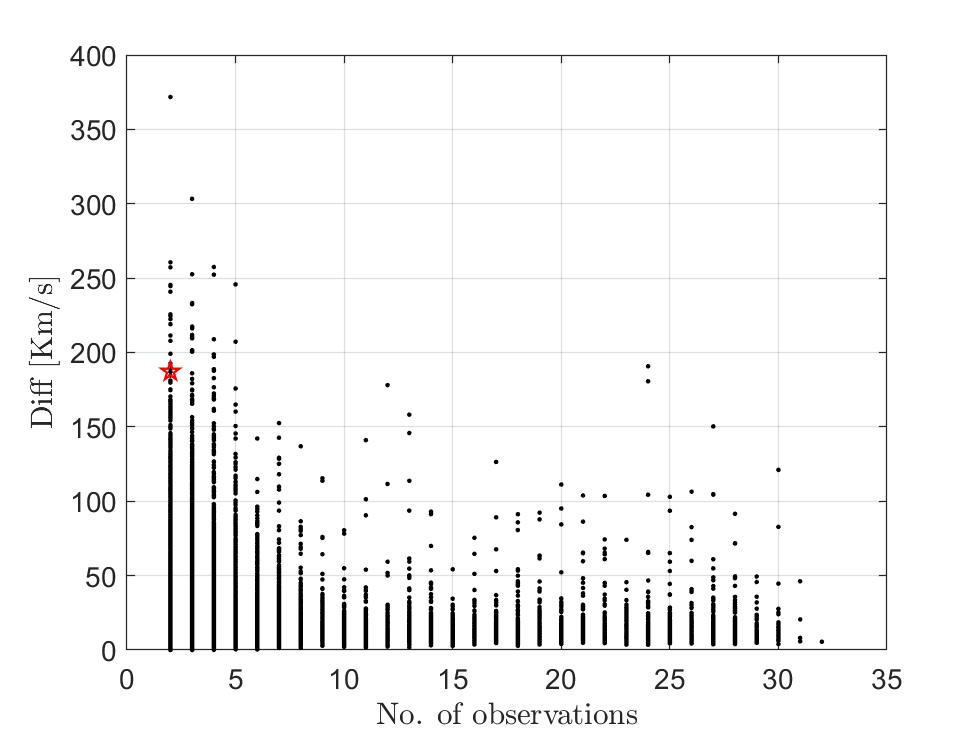}
\caption{RV differences vs.~number of observations in LAMOST DR6. Red star represents J1255.}
\label{fig:RV range}
\end{figure}

Our initial search for compact objects was based on the 
LAMOST \citep{cui2012large} stellar RV 
DR6 database,\footnote{http://dr6.lamost.org/v1/} which includes on the order of $10^6$ sources with at least two measurement epochs. 
We searched for stars showing large RV differences between the available measurements, which could indicate binary motion with a massive companion.
The search identified $955\,882$ stars with more than one RV measurement; a histogram of the number of epochs per star is presented in Fig.~\ref{fig:histogram}.
Some of these stars display large RV
differences,
as seen in Fig.~\ref{fig:RV range}. 

\begin{figure} 
\centering
\includegraphics[height=5cm,width=8cm]
{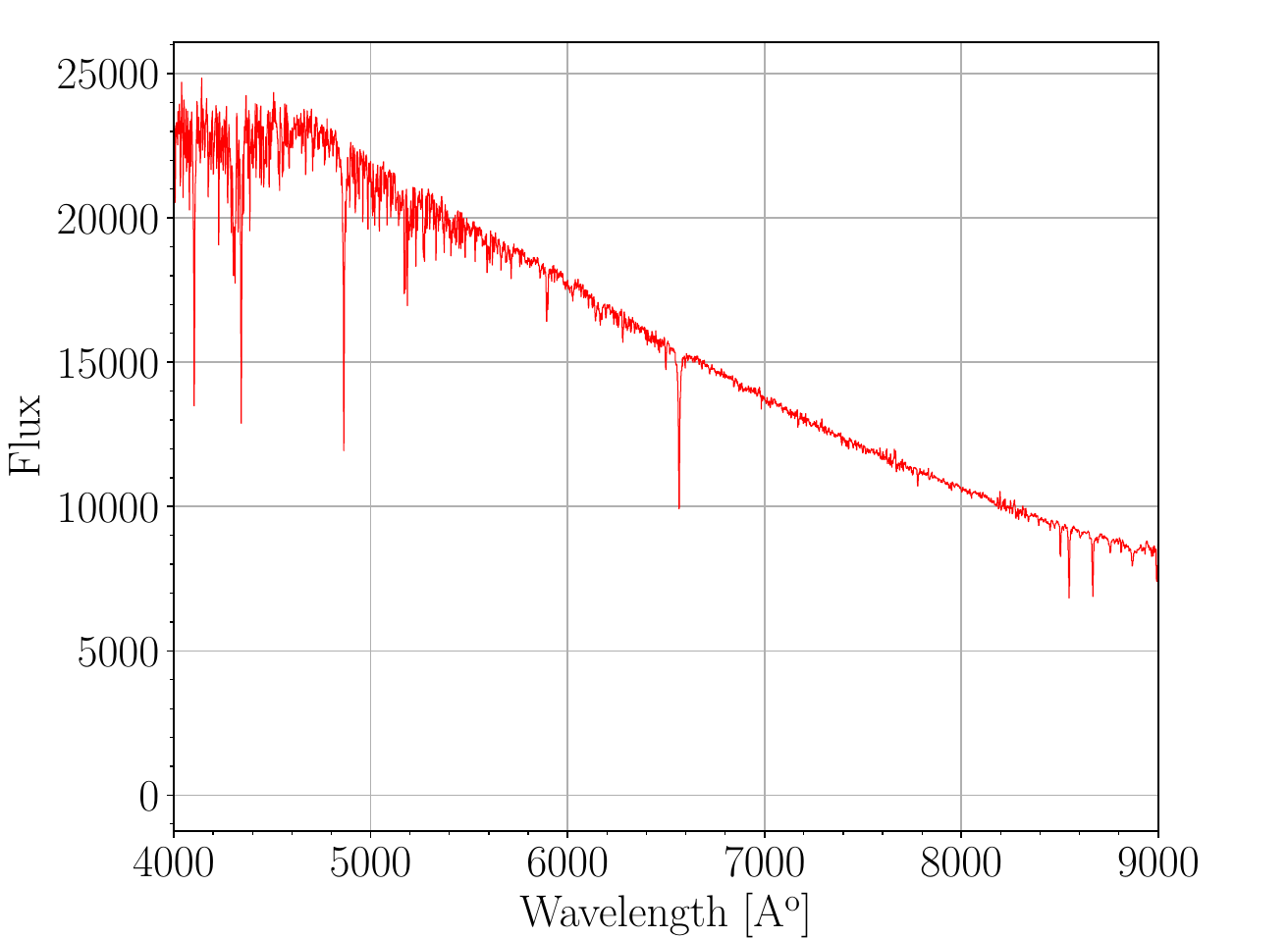}
\caption{LAMOST spectrum of J1255. Flux in arbitrary units.}
\label{fig:LAMOST Spectrum}
\end{figure}

One such, 
relatively bright (V=11.9)
star --- 
TIC (\textit{TESS} Input Catalog) 150388416 
\citep[][]{stassun19},\footnote{https://tess.mit.edu/science/tess-input-catalogue/}
the faint X-ray source 2XMM~J125556.57+565846.4 \citep{2XMM_09,4XMM_20} (hereafter J1255), 

has two LAMOST measurements separated by 185 km s$^{-1}$, as seen in the figure, suggesting a large-amplitude RV modulation. The LAMOST spectrum of J1255,  shown in Fig.~\ref{fig:LAMOST Spectrum}, indicates a normal F-type star. 

The \textit{TESS} photometry (see Section~\ref{sec:TESS} below) indicated the star might be a good candidate for our search for dormant compact objects.

\newpage
\subsection{TRES and CARMENES  radial velocities}

To derive the binary period and the RV amplitude, we monitored J1255 with two spectroscopic facilities.
We obtained $14$  spectra with TRES\footnote{http://tdc-www.harvard.edu/instruments/tres/} --- the Tillinghast Reflector Echelle Spectrograph,\footnote{https://exofop.ipac.caltech.edu/tess/TRESnote.pdf}
%
%
mounted on the 1.5 m telescope at the Fred Lawrence
Whipple Observatory, Arizona. TRES is a fiber-fed
echelle spectrograph with wavelength coverage from
$3900$ to $9100$ \AA, spanning 51 echelle orders at a resolution of $R \simeq 44000$. 
Wavelength solutions are provided
by ThAr-lamp exposures that bracket each observation. The
observations are reduced as per \citet{buchhave10}. RVs were obtained by correlating the 
orders that contain enough spectral information
with the observed spectrum 
that has
the highest Signal-to-Noise-Ratio (SNR).
The cross-correlation function (CCF) was calculated independently for each spectral order. The RVs and their uncertainties were derived from the combined CCF (see  \citealt{zucker03}).
%

\begin{figure*} 
\centering
\includegraphics[width=\textwidth]
{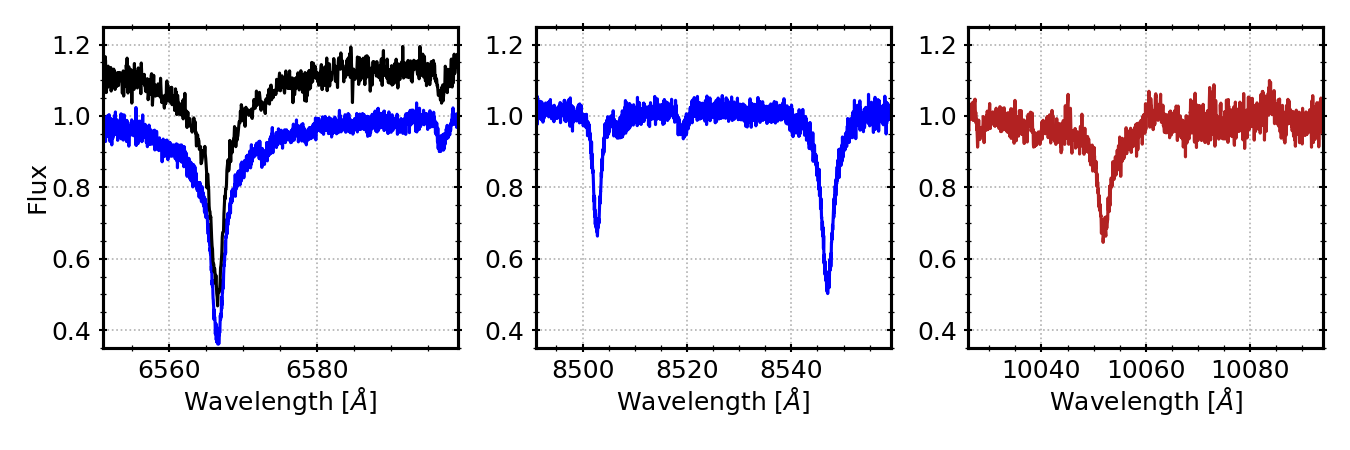}
\caption{TRES (black), CARMENES-VIS (blue) and CARMENES-NIR (red) coadded spectra. The spectra include the H${\alpha}$ line (left),  two of the  Ca\,II  triplet 
lines (middle), and the Pa-$\delta$ line (right). 
}
\label{fig:H_alpha}
\end{figure*}

Five additional spectra were obtained in February 2021 with 
CARMENES\footnote{https://carmenes.caha.es/ext/instrument/index.html} \citep{Quirrenbach2016,Quirrenbach2018}, attached to the Cassegrain focus of the 3.5 m telescope at Calar Alto observatory, Almer\'ia, Spain. CARMENES consists of a pair of cross-dispersed echelle spectrographs, the visual channel (VIS) covering a wavelength range from 5200 to 9200\,{\AA} with  $R\simeq 94600$, and the near-infrared channel (NIR), which covers a wavelength range from 9600 to 17100 {\AA} at $R\simeq 80400$. Both channels use Fabry-P\'erot etalons and hollow-cathode emission-line lamps for wavelength calibration \citep{Schafer2018}. 
The data reduction of CARMENES observations is performed by the \texttt{caracal} pipeline \citep{Caballero2016}, which corrects for bias, flat-fielding, and cosmic ray events, performs wavelength calibration, and extracts one-dimensional spectra. The spectra were corrected for telluric contamination following the procedure described by \citep{Nagel2022}.  Fig.~\ref{fig:H_alpha} shows three of the spectral ranges covered by CARMENES. The panel with H$\alpha$ shows also the TRES spectrum in that range.

\begin{figure}
\centering
\includegraphics[width=0.45\textwidth]
{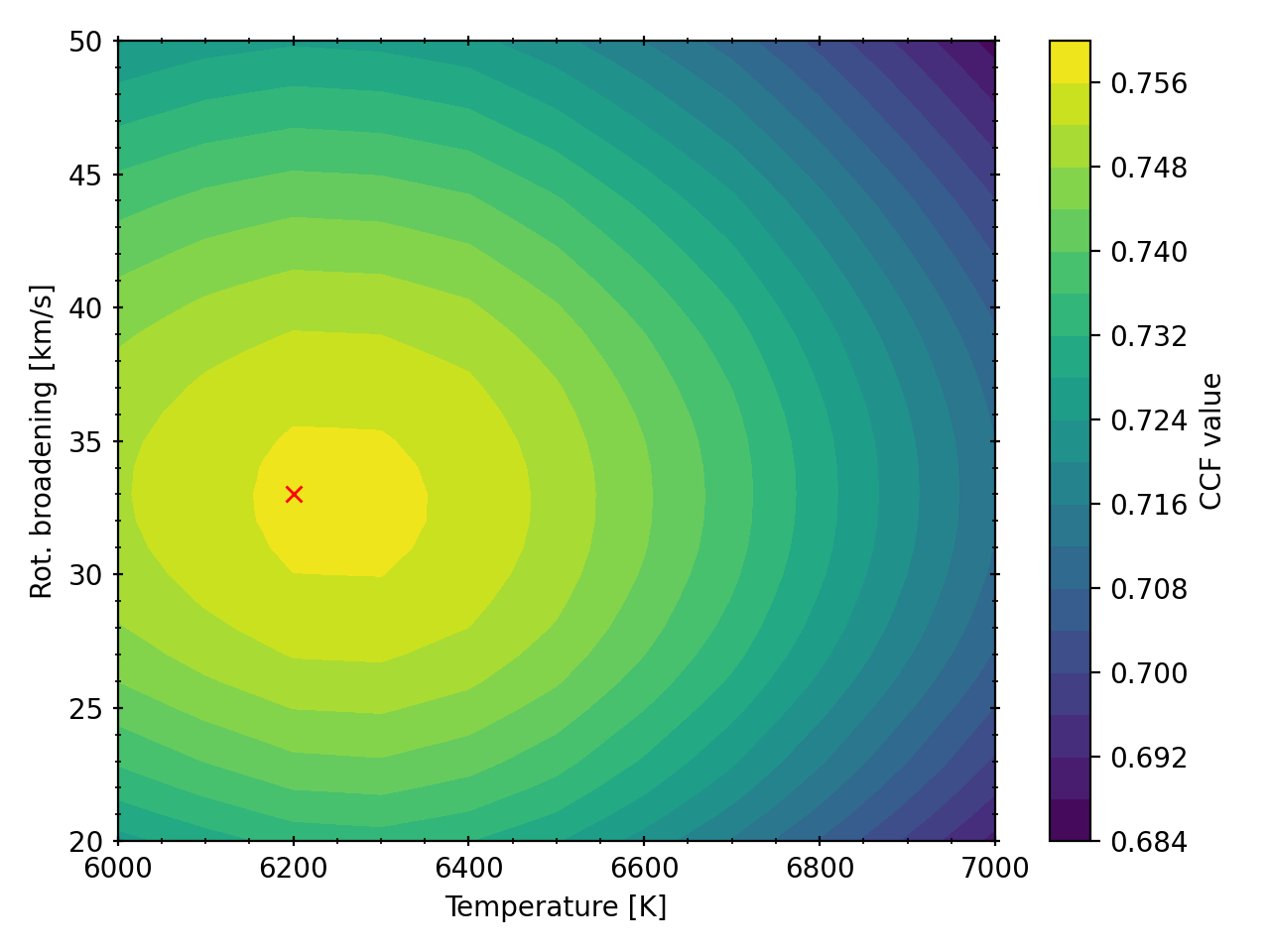}
\caption{Search for the best template to derive the RVs of the CARMENES spectra performed with \texttt{TODMOR}.
The color code indicates the peak value of the cross-correlation function, obtained for different stellar rotations and temperatures, with the red cross indicating the best values. Note the shallow peak along the rotation axis.}
\label{fig:todcorgrid}
\end{figure}

To analyze the 
optical
CARMENES spectra, we first adopted a theoretical template from the library of PHOENIX stellar models \citep{Husser2013} that best fit the observed spectral features.  We explored a grid of values for effective temperature and spectral line broadening (approximately equivalent to the projected rotation velocity, 
${\rm v}\sin i_{\rm rot}$), where $i_{\rm rot}$ is the inclination of the stellar rotation axis, looking for a combination that maximized the cross correlation, as illustrated by Fig.~\ref{fig:todcorgrid}. 
The best-fitting template parameters and $1\sigma$  uncertainties are 
$T_{\rm eff}=6280 \pm60$ K and 
${\rm v}\sin i_{\rm rot}=32.3\pm0.5$ km s$^{-1}$. 
A rotational broadening profile (\citealt{gray92}) was applied to account for the stellar rotation while adopting a limb-darkening coefficient from \citet{claret11}.
This template was then used to derive the CARMENES RVs. 

\subsection{Orbital solution}
\label{sec:orbital}

The RVs measured by the three instruments 
--- LAMOST, TRES and CARMENES ---
are listed in Table~\ref{tab:RVs},
together with their $1\sigma$ uncertainties.
The TRES velocities are relative to the RV at BJD $2459587$. 

\begin{figure} 
\centering
\includegraphics[width=0.45\textwidth]{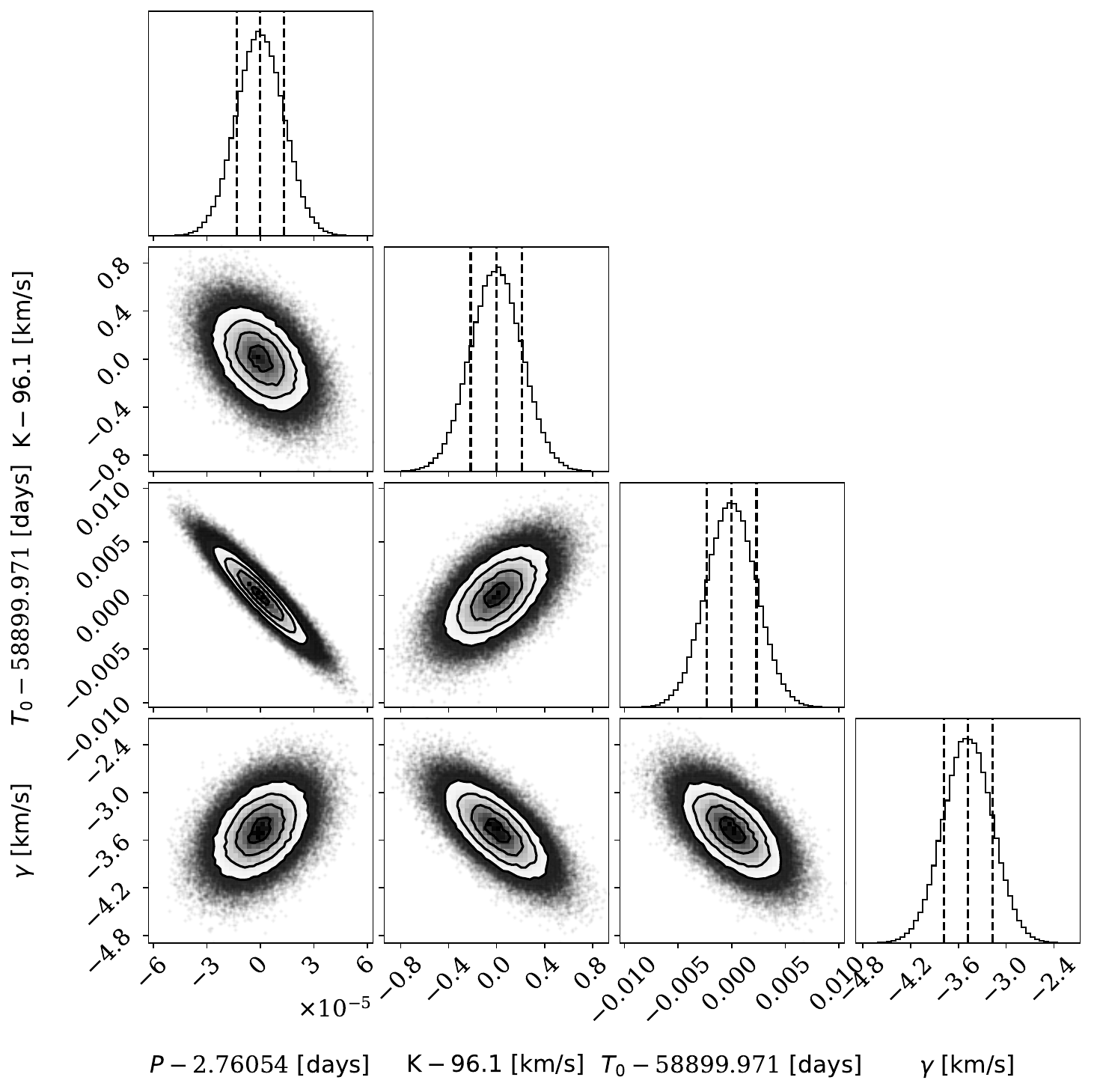}
\caption{MCMC posterior distributions of the orbital parameters --- period, $P$, Semi-amplitude, $K$, time of inferior conjunction, $T_0$ [BJD-$2400000$] and center-of-mass RV, $\gamma$, based on the CARMANES RV offset. Off-diagonal sub-plots present the joint distributions with contours, and diagonal elements show the probability distribution of each parameter, with percentiles of $16\%$, $50\%$ and $84\%$, denoted by vertical dashed lines.  }
\label{fig:corner}
\end{figure}

\begin{figure*} 
\centering
\includegraphics[width=18cm]{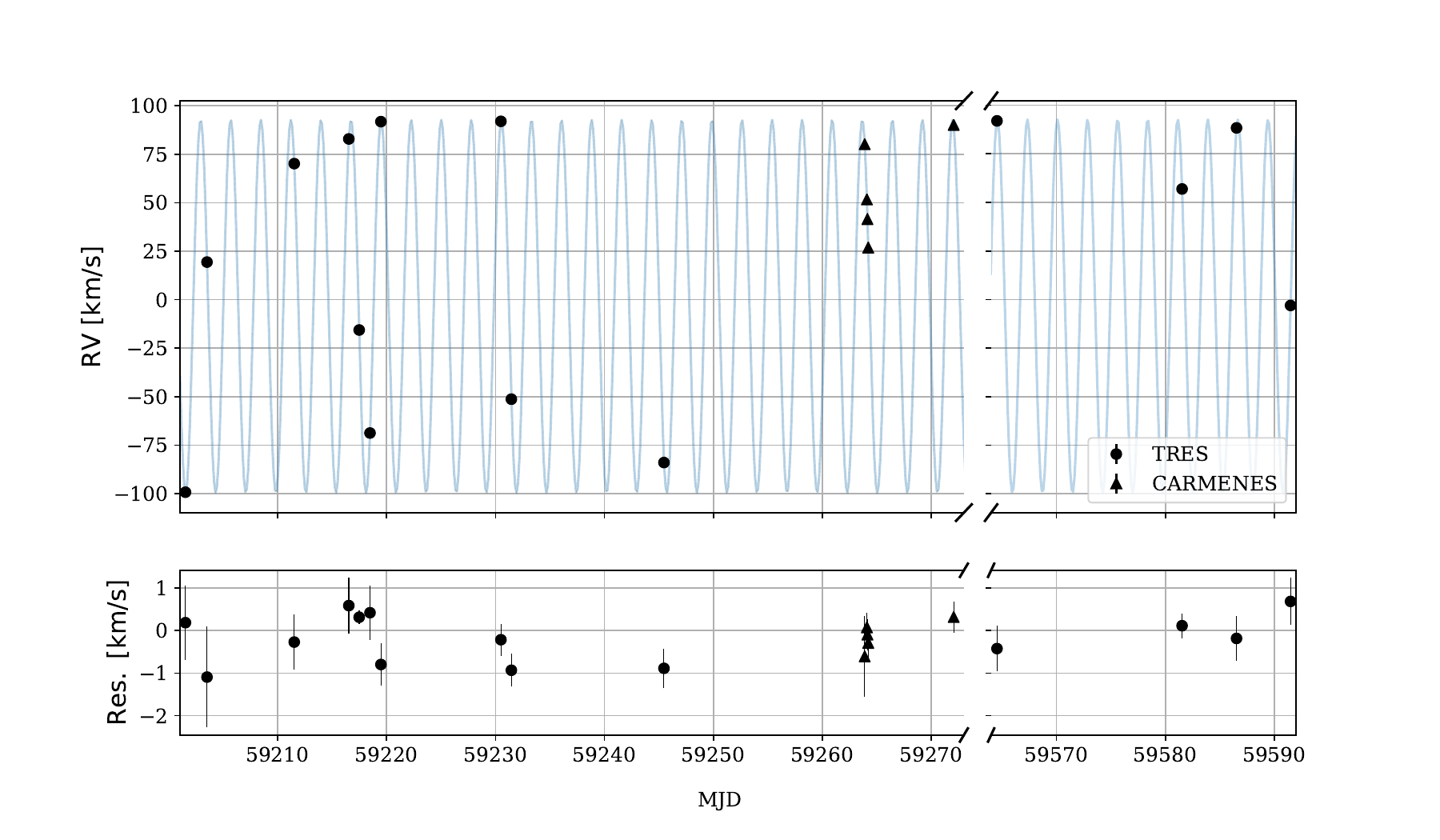}
\caption{Orbital solution for J1255. {\it Top} --- best model with the RV measurements. For display purposes,  we have excluded LAMOST measurements, because of their large time span. 
{\it Bottom} --- residuals from the model.}
\label{fig:ModelTime}
\end{figure*}

\begin{figure*} 
\centering
\includegraphics
[width=\textwidth]
{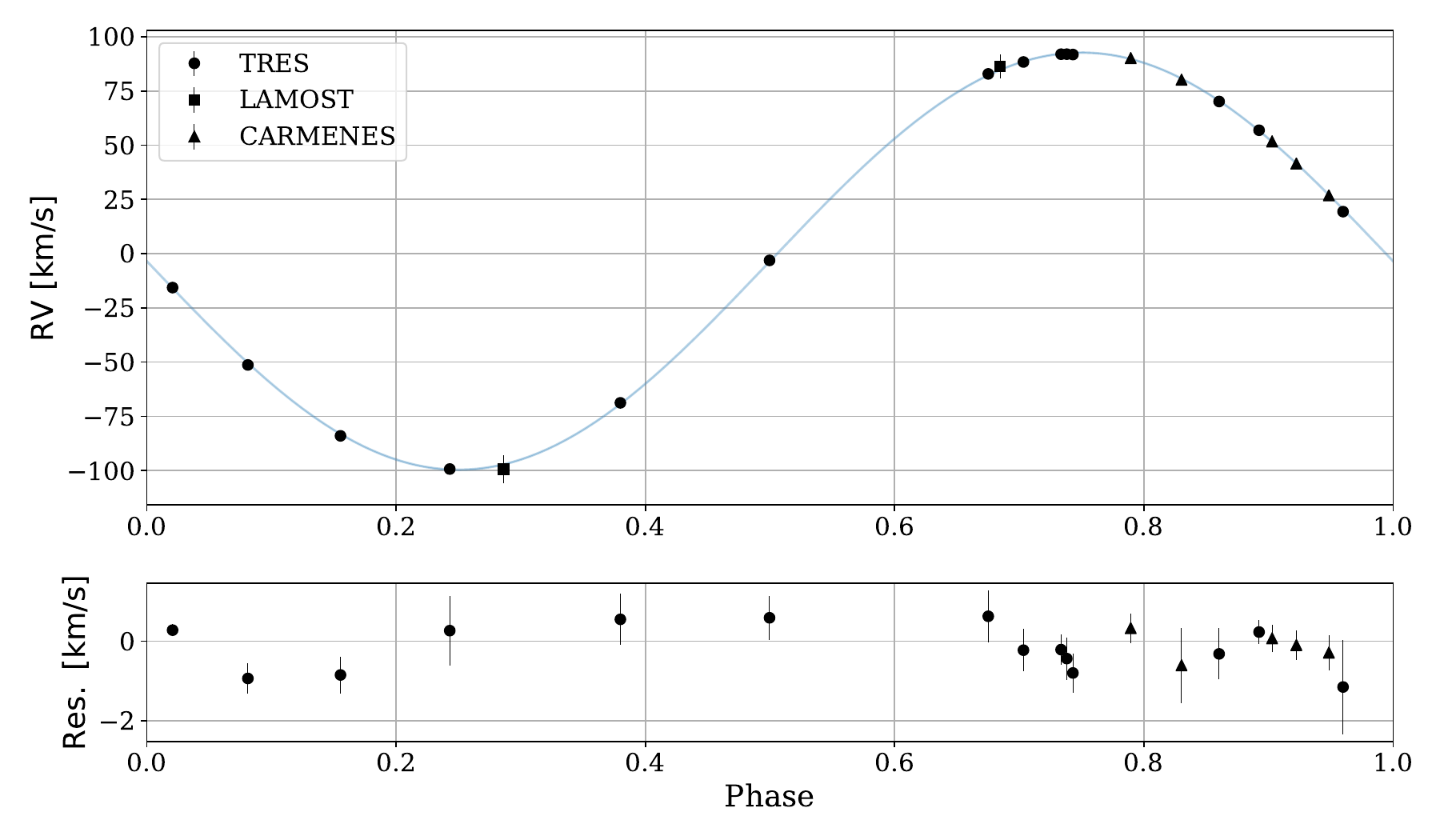}
\caption{
{\it Top} --- radial velocities of J1255, derived from the TRES, CARMENES, and LAMOST spectra, together with the orbital solution, 
folded with the  P=2.760549 d orbital period. 
{\it Bottom} --- residuals. 
}
\label{fig:RV}
\end{figure*}

\begin{table}
\caption{RVs of J1255. 
\label{tab:RVs}}
\begin{center}
\begin{tabular}[t]{lrcc}
\hline
&  & RV & \\
Time & RV \ \ \ \ & Uncertainty & Instrument\tnote{*}\\
BJD-$2450000$&$\mathrm{km~ s}^{-1}$&$\mathrm{km~ s}^{-1}$&\\
\hline
6424.556& -99.94& 6.47& L\\
7474.662&  85.77& 5.42& L\\
9202.041& -187.61&     0.87& T\\
9204.020&  -68.98&     1.18& T\\  
9212.027&  -18.21&     0.64& T\\   
9217.036&   -5.49&     0.65& T\\   
9217.990& -103.99&     0.16& T\\   
9218.982& -157.11&     0.64& T\\   
9219.985&    3.43&     0.49& T\\   
9231.001&    3.60&     0.37& T\\    
9231.960& -139.65&     0.38& T\\   
9245.968& -172.33&     0.46& T\\
9264.394& 80.09&  0.94& C\\
9264.594& 51.66&  0.34& C\\
9264.648& 41.47&  0.37& C\\
9264.720& 26.81&  0.44& C\\
9272.563& 90.05&  0.37& C\\
9565.039&    3.62&     0.53& T\\   
9582.028&  -31.46&     0.29& T\\   
9587.027&    0.00&     0.53& T\\   
9591.986&  -91.45&     0.56& T\\
\hline
\end{tabular}
     \begin{tablenotes}
       \item [*] L-LAMOST, T-TRES, C-CARMENES
     \end{tablenotes}
  \end{center}
\end{table}
To derive an orbital solution for J1255, we applied the \texttt{exoplanet} package \citep{foreman2021} with the probabilistic model in {\tt  PyMC3} \citep{salvatier2016},  
allowing for different RV offsets of the three spectrographs. 
We first explored a non-zero eccentricity model, finding insignificant eccentricity  at  a 2$\sigma$ limit of  $e<8 \times 10^{-3}$. We therefore adopted a circular model, 
 (zero eccentricity and zero longitude of the periastron), with an
uninformative prior for the semi-amplitude, $K$, orbital period, $P$, the time of inferior conjunction, $T_0$, and $\gamma$, the RV of the center of mass of the system. 
Posterior distributions of the fitted parameters are shown in Fig.~\ref{fig:corner}, and the adopted parameters are given in Table~\ref{tab:RVsolution}. 
The best-fit model, together with the data and residuals, are plotted in Fig.~\ref{fig:ModelTime} and \ref{fig:RV}. 

To better constrain the possibility of a long-term RV trend, we attempted to include a linear trend in the solution. Based on the TRES dataset, we constrain the slope to $-0.0015 < a < 0.0030$ km~s$^{-1}$/day at the $2\sigma$ level, indicating that any such a linear trend is insignificant, within the range accessible to our precision and data time span. 

\begin{table}
\caption{Orbital solution for J1255. 
\label{tab:RVsolution}}
\begin{center}
\begin{tabular}{ c c r r c }
      & $P$      & $T_0$ \ \ \ \ \ \ \  &$K$\ \ \ \ & $\gamma$  \\
      &[day]    &  [BJD] \ \ \ &  [km~s$^{-1}$] \    &   [km~s$^{-1}$]  \\
\hline
      &2.760541 & 2458899.970 &   96.1  & \, -3.4   \\ 
$\pm$ &0.000013 &   0.002    &   0.2 & 0.3    \\
\end{tabular}
\end{center}
\end{table}
\newpage

Just before submission of the paper, the {\it Gaia} orbit of J1255, Gaia DR3 1577114915964797184, became available. Their elements --- period of $2.760586\pm  4.9\times 10^{-5}$\,d, 
semi-amplitude
of $96.2 \pm 1.3$\,km s$^{-1}$ and eccentricity of 
$0.019 \pm 0.011$ \citep{NSS}, are consistent with our spectroscopic orbit. 

\section{Photometric modulation of J1255 --- \textit{TESS} and UVOT}
\label{sec:photometry}

\subsection{ \textit{TESS}}
\label{sec:TESS}
\begin{figure*} 
\centering
\includegraphics[width=\textwidth]
{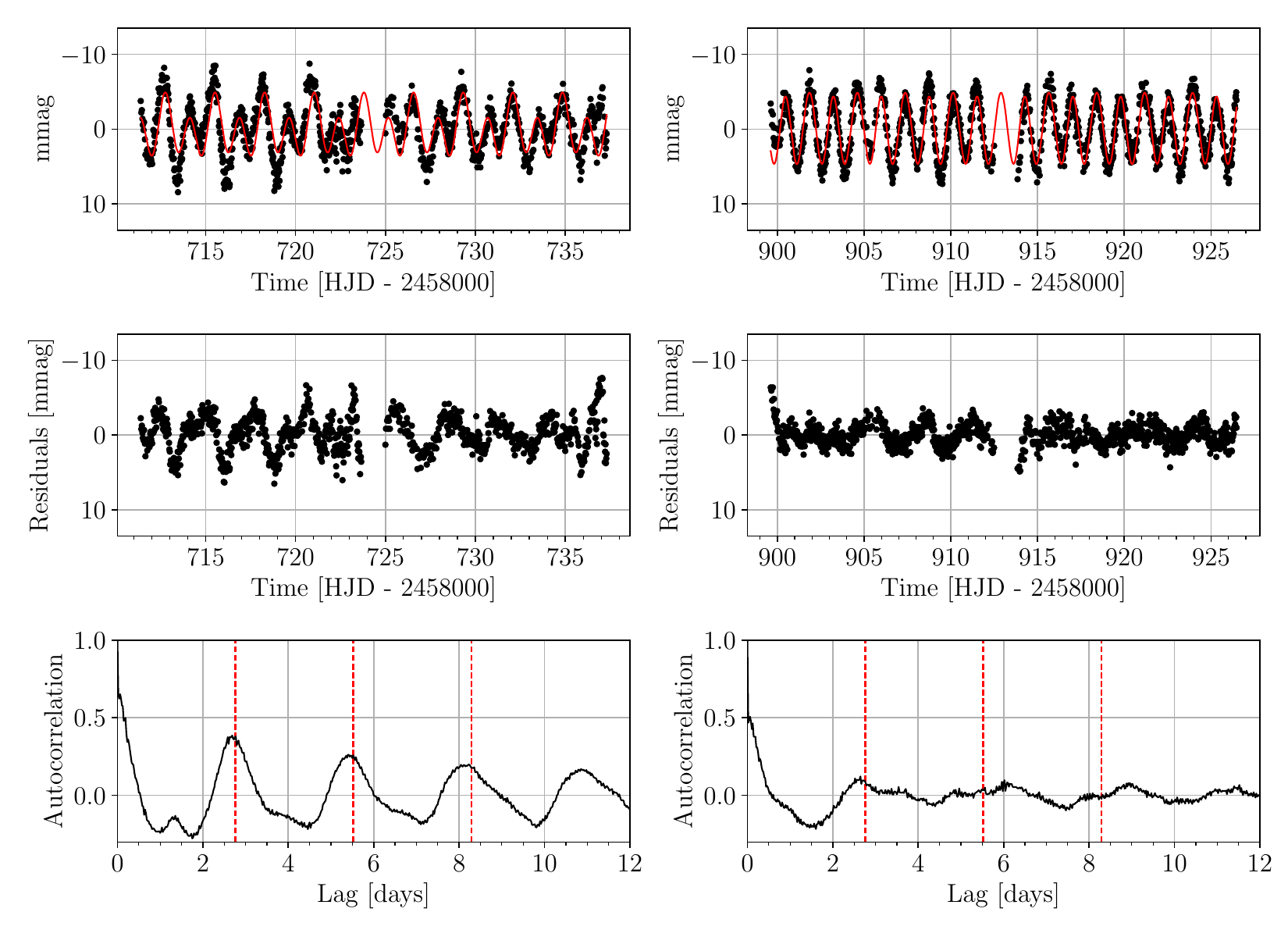}
\caption{{\it TESS} photometry obtained from Sectors 15 (left) and 22 (right).
{\it Top} --- light curve in mmag (black), after removal of long-term trend, and a three-harmonic model applied to that sector (red). 
Orbital period and phase were held fixed at the RV-derived values.
{\it Middle} ---  residuals from the models.
{\it Bottom} --- autocorrelation function of the residuals, where 
red, dashed, vertical lines mark 1$\times$, 2$\times$ and 3$\times$ the orbital period.
}
\label{fig:TESS-unfolded}
\end{figure*}

J1255 was observed by \textit{TESS} --- the Transiting Exoplanet Survey Satellite \citep{TESS15} during Sectors~15 and 22.
Each sector includes observations for $27.4$ days, with full-frame images recorded continuously at a cadence of $\sim 30$\,min.
A 1.5-day gap at the midpoint of each sector is a result of the data downlink from the satellite.
The light curve of J1255 was extracted from the full-frame images using the \texttt{eleanor} package \citep{feinstein19}, and is plotted in the upper panels of  Fig.~\ref{fig:TESS-unfolded}.

The \textit{TESS} variability is dominated by a periodic signal at half the orbital period, consistent with an ellipsoidal modulation caused by the tidal distortion of the primary star induced by the gravity of the unseen companion \citep{kopal59,morris85,morris93}. 
This variability, at the level of $5$\,ppt (parts per thousands of the average stellar luminosity), is plotted and modeled in Fig.~\ref{fig:TESS-unfolded} using the first three harmonics of the orbital period and the phase derived from the RVs in Section \ref{sec:orbital}.
Low-frequency systematic effects were filtered using a Fourier comb, applied at frequencies lower than $0.25\times$ the orbital frequency, simultaneous with the harmonic fit \citep{faigler11}.

Fig.~\ref{fig:TESS-unfolded} also shows the residuals from the fit, which display some additional variability at a level of $2$--$3$\,ppt, most clearly in the data from Sector~15.
The extra variability, displayed in the middle panels of the figure, is not strictly periodic, and therefore we opted to present its autocorrelation (ACF), as in \citet*{mcquillan13}. The ACF shows that the variability is indeed strongly correlated
with itself at earlier times,
at a typical lag that is similar to the orbital period.  This 
probably originates from stellar rotation, as observed in many stars by recent space missions \citep{mcquillan14}.
The stellar rotation period is 
equal or almost  equal to the orbital period, apparently
because stellar rotation 
tends to be synchronized
with orbital period at such short periods \citep[e.g.,][]{mazeh08}.

\begin{figure} 
\centering
\includegraphics[scale=0.45]
{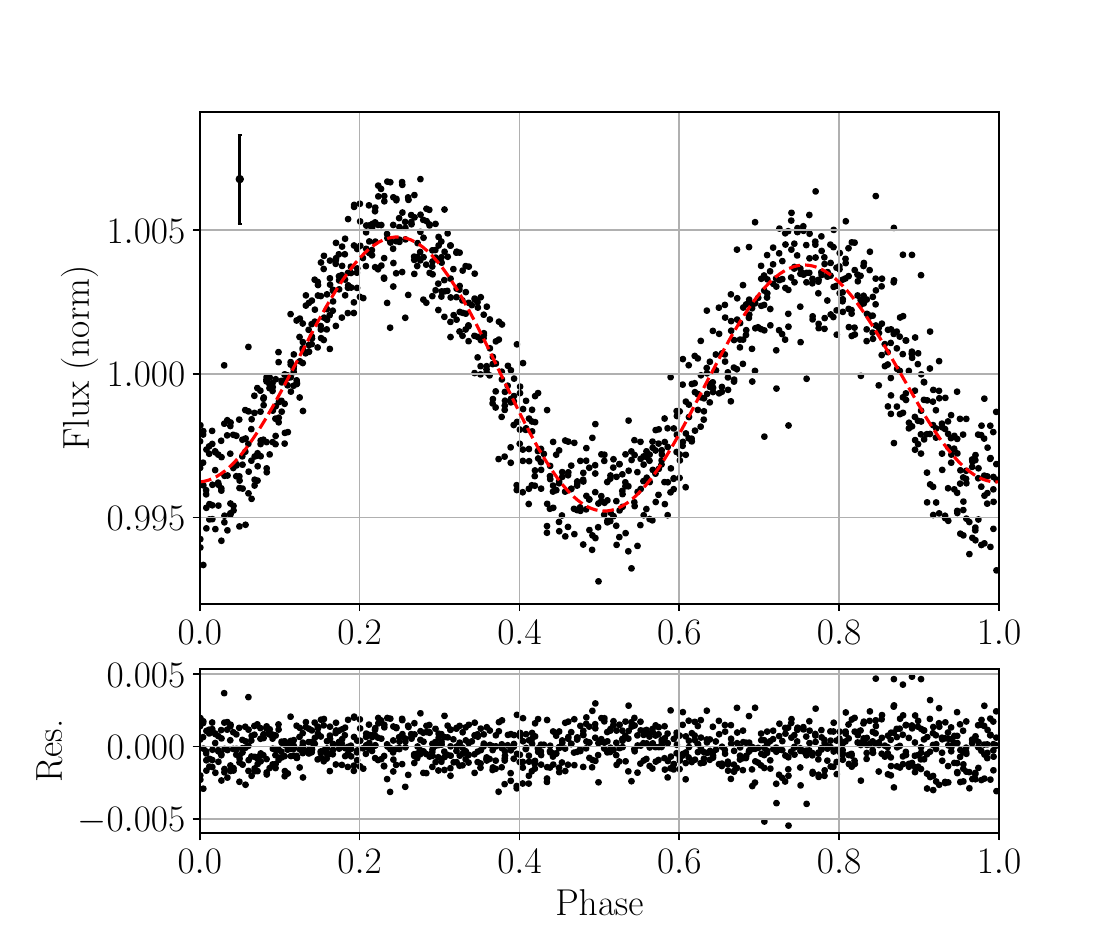}
\caption{{\it TESS} photometry from Sector 22, 
phase-folded with the RV orbital period, 
showing the ellipsoidal modulation of the F-type star. {\it Top} --- 
TESS data points are in black. Red line is the best-fit two-harmonic model. The typical uncertainty of the data points is plotted in the upper left corner. The ellipsoidal modulation is at half the orbital period. {\it Bottom} --- residuals from the model.}
\label{fig:TESS-folded}
\end{figure}

For our canonical measurement of the ellipsoidal amplitude, we use only Sector~22, in which the residuals are significantly smaller and less correlated.
The phase-folded data from Sector 22 are shown in Fig.~\ref{fig:TESS-folded}, along with the best-fit two-harmonic model. 
As before, the period and phase of the fit were fixed at the RV orbital-solution values.

The cosine and sine amplitudes of the two harmonics
($a_{1c}$,$a_{1s}$,$a_{2c}$,$a_{2s}$) in units of the median flux of the system
are given in the upper part of Table~\ref{tab:harmonic}. Also given are the total amplitudes of these two harmonics ($A_1$ and $A_2$).
The ellipsoidal amplitude of the best-fit model is $4.27 \pm 0.05$\,ppt.

\subsubsection{Beaming modulation?}

Note that we expect another photometric modulation --- the beaming effect, induced by the radial velocity of the optical star  \citep[e.g.,][]{zucker07,faigler11,BEER15}, which appears as a sine function of the orbital phase  for a circular orbit.
For a sun-like star in the \textit{TESS} band
the amplitude is $\sim4V_{\rm orb}/c$, where $V_{\rm orb}$ is the orbital velocity and $c$ is the light velocity.  In our case the expected amplitude is $\sim1.3$ ppt. The observed modulation, represented by 
the $a_{1s}$ coefficient, is only $\sim0.5$ ppt.

Apparently, the other variability, probably due to the stellar spots, masks 
the full manifestation of the beaming modulation. We probably need a longer time span, like the \textit{Kepler} light curves \citep[e.g.,][]{faigler12,faigler13}, to  average out all other variabilities, in order to obtain the correct beaming amplitude. 

\begin{table}
\caption{Harmonic amplitudes of J1255 in terms of the median flux.
\label{tab:harmonic}}
\begin{center}
\begin{tabular}[t]{lcccccc}
\hline
 & $a_{1c}$  & $a_{1s}$&  $A_{1}$ & $a_{2c}$ & $a_{2s}$& $A_{2}$  \\
  & [ppt]& [ppt]& [ppt]& [ppt]&  [ppt]& [ppt] \\
\hline
TESS & $0.501 $ & $0.494 $& $0.703$ & $-4.266 $ & $-0.086 $ & $4.267 $ \\
\ \ \ \ $\pm$     & $0.054 $ & $0.054 $& $0.054$ & $0.054 $ & $0.054 $ & $0.054 $ \\
UVOT & $2.681 $ & $-5.042 $& $5.71$ & $-4.194 $ & $7.322 $ & $8.438 $\\ 
\ \ \ \ $\pm$ & $1.606 $ & $1.514 $& $1.535$ & $1.532 $ & $1.59 $ & $1.576 $  \\
\hline
\end{tabular}
  \end{center}
\end{table}


\subsection{Swift UVOT photometry}
\label{sec:Swift}

The {\it Neil Gehrels Swift Observatory}\footnote{https://swift.gsfc.nasa.gov/}
\citep{swift04} 
obtained 187 short exposures of the active galaxy Mrk 231 over the course of a year, resulting in serendipitous UV measurements of J1255, which was within the field of view of Swift's UVOT camera in all pointings. Exposure times ranged from $\sim200$~s to $\sim1700$\,s, with a median of $\sim 1000$\,s. 

\begin{figure} 
\centering
\includegraphics[scale=0.5]
{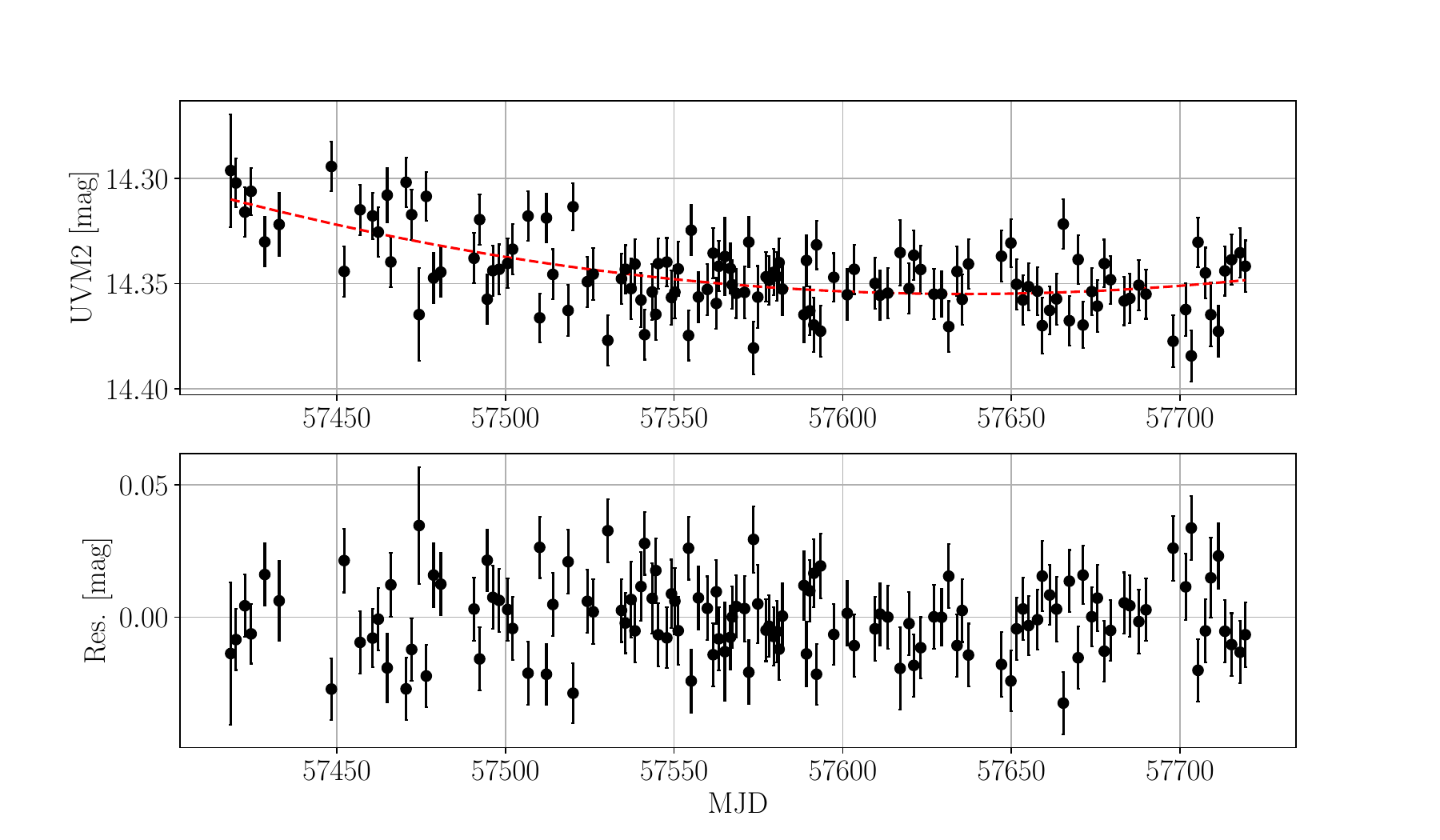}
\caption{UVOT UVM2 light curve 
of J1255. \textit{Top} --- Original photometry (see text), with a fitted second-order polynomial. \textit{Bottom} --- Photometry after the long-term removed.
%
%
}
\label{fig:UVOT}
\end{figure}

The single UVOT frames were downloaded from 
{\sc HEASARC},\footnote{https://heasarc.gsfc.nasa.gov} and light curves were extracted with the {\it photutils} package from {\it astropy} \citep{astropy13}, using the optimal aperture (10’’) and zero-point magnitude for the {\it UVM2} filter given by \cite{poole08}. 
%
%
The resulting light curve,  after outlier removal, is composed of 123 individual epochs, all of
which are plotted in Fig.~\ref{fig:UVOT}.
A long-term trend, probably due to a slow degradation of the detector, is  clearly visible in the data. Similar trend was seen in the intensity of a nearby star. This trend was fitted and removed before a periodic analysis was performed.

\begin{figure}
\centering
\includegraphics[scale=0.45]
{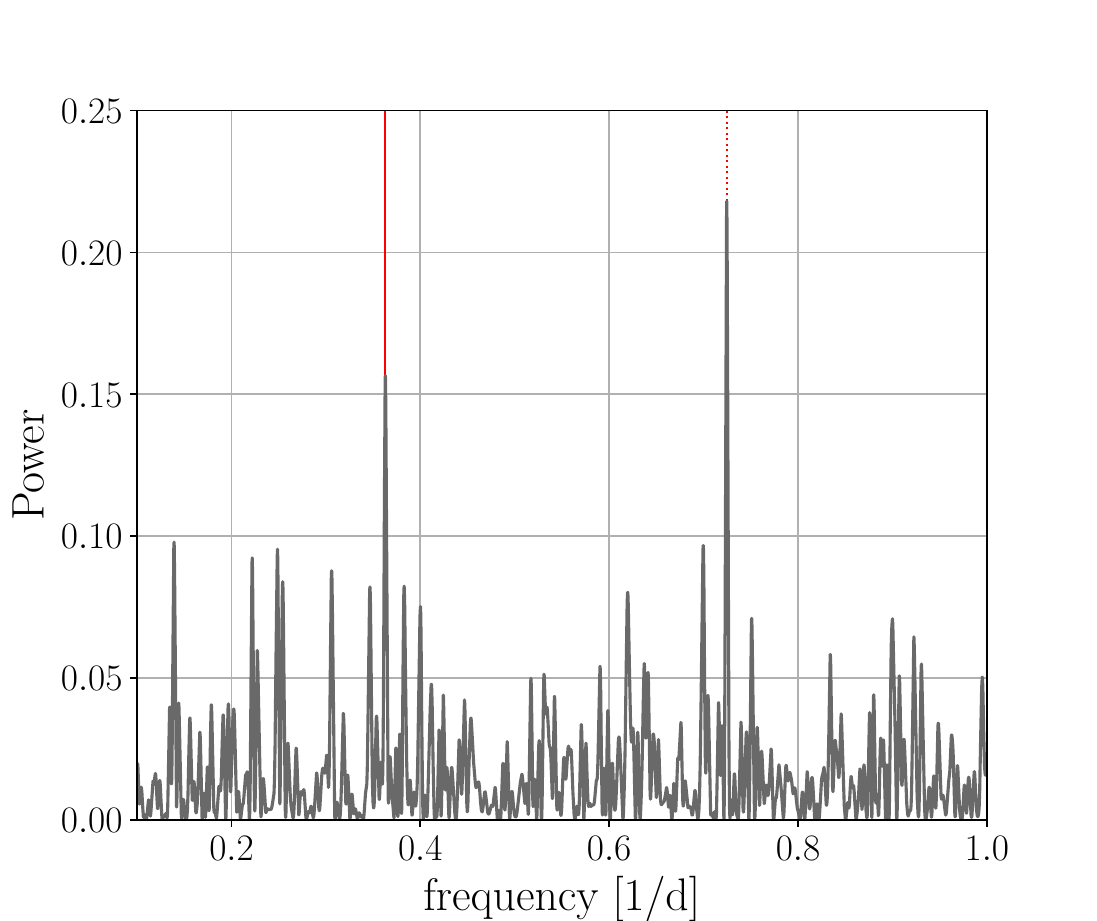}
\includegraphics[scale=0.5]
{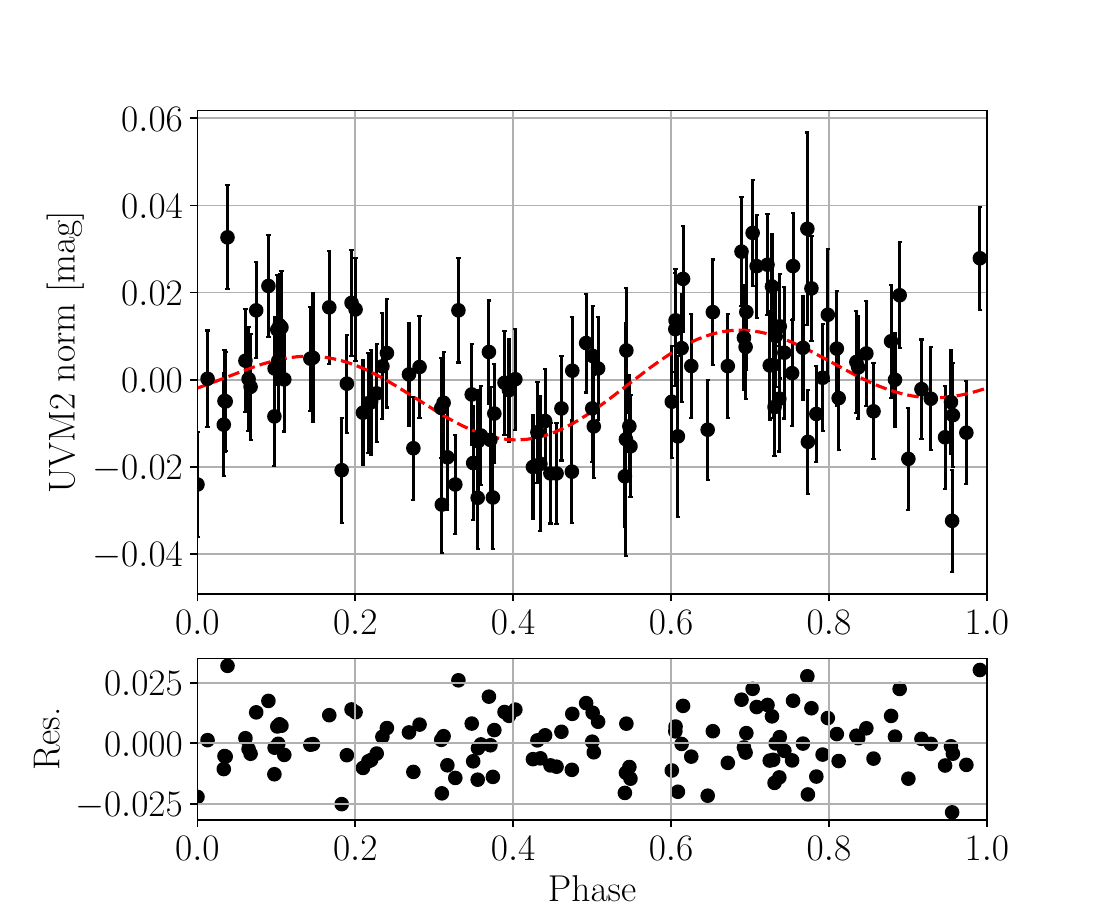}
\caption{Periodogram and phase-folded UVOT UVM2 light curve. {\it Top} --- GLS periodogram of the normalized flux. The solid and dashed vertical red lines represent the full and half orbital period as determined from the RV data. 
 {\it Middle} --- folded data, with a two-harmonic model at the orbital period. 
{\it Bottom} --- residuals from the model.
}
\label{fig:UVOT_folded}
\end{figure}

To search for any periodic signal in the data we
applied the
generalized Lomb-Scargle analysis \citep{Zechmeister2009}. 
The resulting periodogram, shown in the upper panel of Fig.~\ref{fig:UVOT_folded}, exhibits a significant peak at $\mathrm{P}=(1.3799\pm0.0005)$~d, where the period uncertainty was derived by a Monte Carlo approach. This is in good agreement with half the orbital period of the system,
as expected for a system with ellipsoidal modulation. 

To derive the shape and amplitude of the modulation, we fitted a two-harmonic model to the phased data, which was normalized to the average flux in that part of the data. We used the orbital period and phase of the RVs, as was done with the TESS photometry.
The cosine and sine amplitudes of the two harmonics
are given in the lower part of Table~\ref{tab:harmonic}, as was done with the {\it TESS} data.
Also given are the total amplitudes of these two harmonics, $A_1$ and $A_2$.


The shape of the UVM2 modulation, with a deeper minimum at phase $\sim 0.5$ and a lower maximum at $\sim 0.75$, is similar with that of the TESS modulation, with a possible small phase shift; see Section~\ref{sec:TESS}.
The second harmonic amplitude of the UVOT data, $8.4\pm1.6$\,ppt, is larger by a factor of two than that of the {\it TESS} data, although the significance of the difference is only at the $2\sigma$ level. 

The derived periodogram includes on the order of 100 independent frequencies, and therefore the probability of obtaining the highest peak at the expected frequency, based on the RV and {\it TESS} photometry is $\sim 0.01$.
The amplitude we derived is five times larger than its uncertainty, indicating a significance of $5\sigma$. Furthermore, the phase of the modulation is consistent with the optical and RV phase, indicating again that the origin of the UV modulation is similar to that of the optical one.


\subsection{XMM-Newton observations of J1255}

\begin{figure} 
\centering
\includegraphics[scale=0.25]
{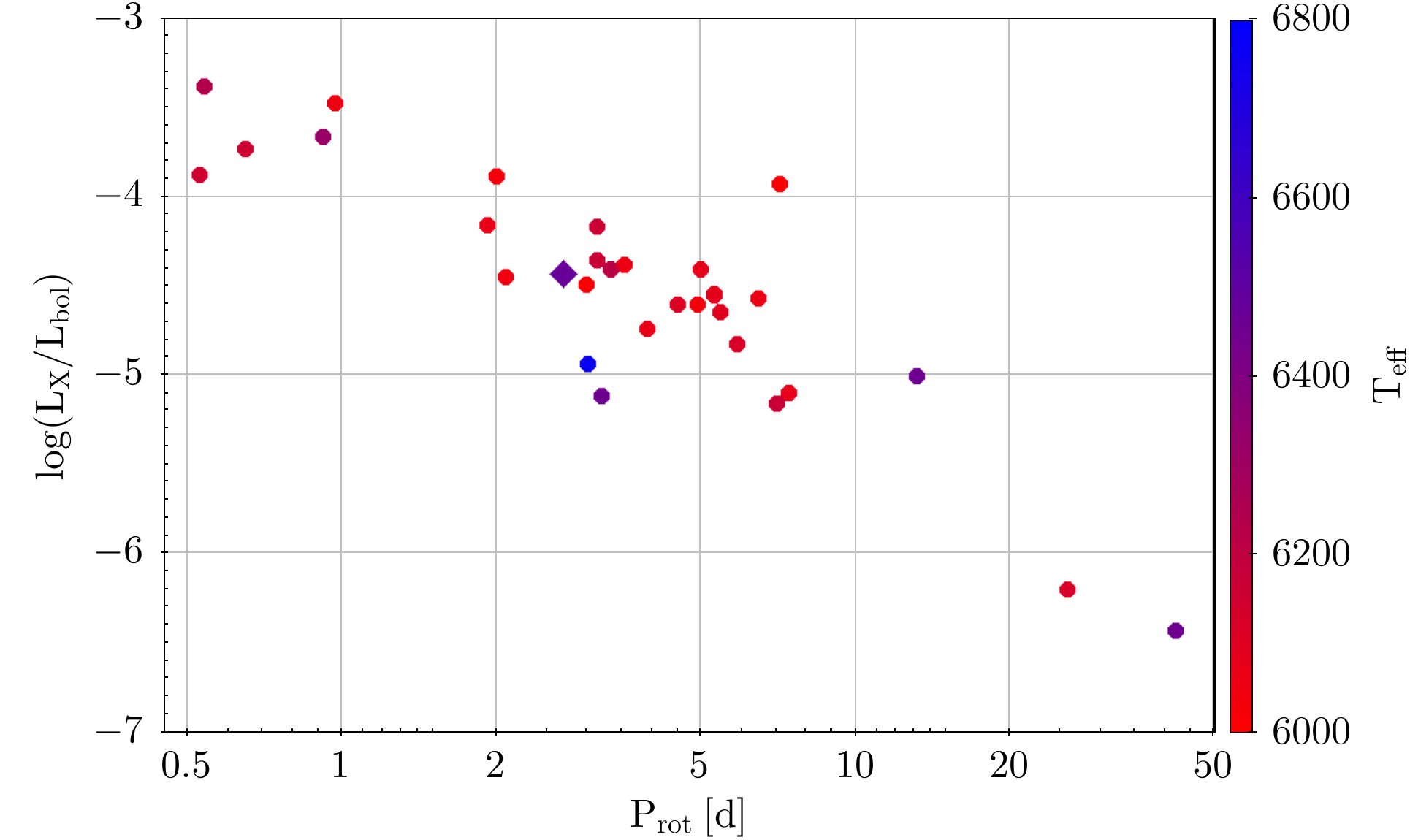}
\caption{Rotation--X-ray activity relation for F-type stars. The ratio of X-ray to bolometric luminosity vs. ~rotation periods for 30 stars with similar effective temperatures $6000\,\mathrm{K}\leq\mathrm{T}_\mathrm{eff}\leq6800\,\mathrm{K}$, as denoted by their color coding, are plotted from the catalog of \citet{wright2011}. J1255 appears as a diamond.}
\label{fig:rot_act}
\end{figure}

J1255 was also detected in  serendipitously  X-rays during three XMM-Newton observations of the galaxy Mrk 231 (observation IDs 0081340201, 0770580401 and 0770580501)  \citep{Franceschini2003,Reynolds2017},  as with the galaxy's {\it Swift} UVOT observations
(Section \ref{sec:Swift}). To check for the presence of any X-ray emission in the  $0.2$--$12$ keV range that is in excess of that expected from the F star (e.g., from accretion on to a companion compact object), we derived the X-ray luminosity for each of the observations as found by \citet{traulsen2020}, yielding an average of $\mathrm{F}_\mathrm{X}=(1.49\pm0.31)\cdot10^{-14}$~mW/m$^2$, corresponding to $\mathrm{L}_{\mathrm{X}}=(6.83\pm1.41)\cdot10^{29}$~erg/s, assuming the parallax from the \textit{Gaia} EDR3 catalog \citep{browngaia2021}. 
With $\mathrm{L}_\mathrm{bol}=4.93\,\mathrm{L}_{\odot}$ from \textit{Gaia}, and scaling to the distance of the target, we get $\mathrm{log(}\mathrm{L}_\mathrm{X}/\mathrm{L}_\mathrm{bol}\mathrm{)}=-4.46\pm0.07$. 

We compare the X-ray-to-bolometric luminosity ratio with the rotational periods and $\mathrm{L}_\mathrm{X}/\mathrm{L}_\mathrm{bol}$ of stars with similar effective temperatures from the catalog published by \citet{wright2011}, assuming the star is tidally locked, and hence has a rotation period equal to the orbital period of the system. The resulting rotation-activity relation, plotted in Fig.~\ref{fig:rot_act}, shows that J1255 has an X-ray emission that matches that of MS stars of the same spectral type and similar rotation period. We can thus conclude that J1255 does not display any significant excess X-ray emission.

\section{Stellar parameters of the primary Star}
\label{sec:stellar_parameters}

In this section we estimate the mass, radius, metallicity and age of the J1255 primary F star.

\subsection{Atmospheric parameters of J1255}
\label{sec:J1255_SED}

\begin{table}
\caption{Atmospheric parameters for J1255
\label{tab:atmospheric}}
\begin{center}
\begin{tabular}{ l r r r r}
\hline
& $T_{\rm 1, eff}$& $\log g$\ \ & [Fe/H]  
& ${\rm v}\sin i_{\rm rot}$\\
&[K]          &[c.g.s] &   & [km s$^{-1}]$ \\
\hline
LAMOST  &   6210& 4.10& $-$0.49 
&   \\
\quad \quad $\pm$   &    15 & 0.02& 0.01 
& \\
\hline
TRES    &   6165& 4.30& 0.06  
& 34.6\\
\quad \quad$\pm$   &    50 & 0.10& 0.08  
& 0.6\\
\hline
CARMENES           &   6280&    &   
& 32.3\\
\quad \quad$\pm$   &    60 &    &   
& 0.5\\
\hline
 TIC        & 6200 & 3.95 & -0.48\\
\quad \quad $\pm$       & 140  & 0.09 & 0.01 \\
\hline
\hline
\end{tabular}
\end{center}
\end{table}

\begin{figure} 
\centering
\includegraphics[scale=0.34]
{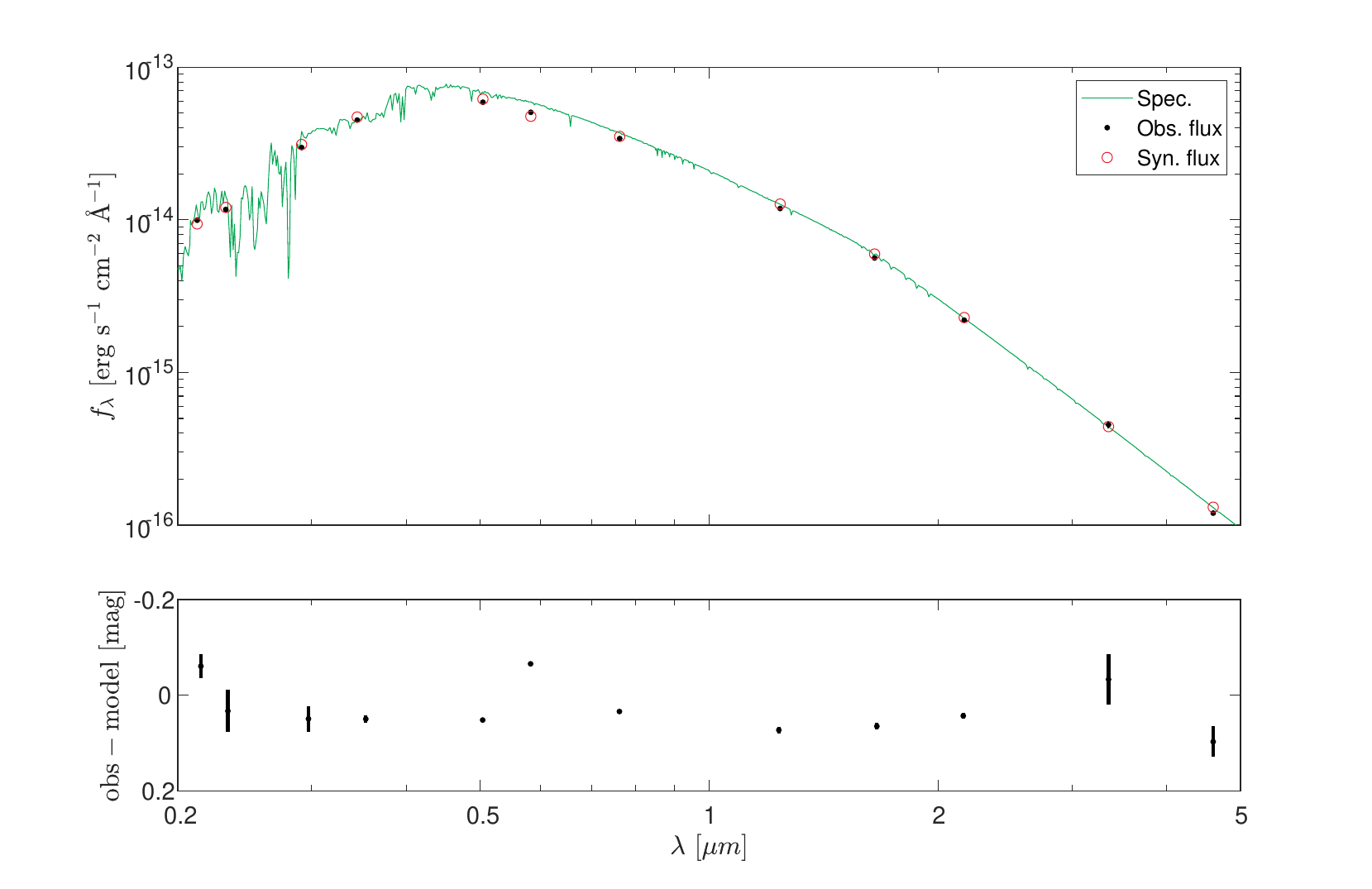}
\caption{Spectral Energy Distribution of J1255. 
{\it Top} --- black points are the measured average photometry (from left to right) for the XMM (UVW1, UVM2, UVW2, U), Gaia (BP, G, RP), 
2MASS (J, H, ${\rm K_s}$)
and WISE (W1, W2) bands. Synthetic spectrum and photometry are plotted with a green line and small red circles, respectively, using the atmospheric models of \citet{castelli03} with  $T_{\rm eff}=6250$\,K, $\log g=4$, and metallicity of $-0.5$. 
We adopted here a stellar radius of 
$R_*=1.9\, R_{\odot}$ at a distance of 600 pc.
{\it Bottom} ---  photometric residuals, for each filter, in mag.}
\label{fig:J1255_SED}
\end{figure}

Table~\ref{tab:atmospheric} lists the atmospheric parameters of J1255, as derived from the TRES and LAMOST spectra, as well as the temperature from TIC. 
Fig.~\ref{fig:J1255_SED}  compares  the consistency of the LAMOST parameters with the broad-band photometric spectral energy distribution (SED) of J1255,  
as collected from Vizier,
given here in mag.\footnote{http://vizier.unistra.fr/vizier/sed/} 
It includes
photometry from \textit{Gaia}: 
${\rm BP} = 12.131 \pm 0.020$, 
${\rm G} = 11.877 \pm 0.020$, 
${\rm RP} = 11.468 \pm 0.020$;
2MASS:
${\rm J} = 11.034 \pm 0.021$, 
${\rm H} = 10.781 \pm 0.023$, 
${\rm K_s} = 10.732 \pm 0.020$;
and WISE:
${\rm W_1} = 10.685 \pm 0.023$, 
${\rm W_2} = 10.696 \pm 0.020$; in which we adopted a minimum uncertainty of $0.02$ mag. 
We plot the original wavelength and flux of an observed photometric band for filters having  a single measurement, but replace them with weighted-average fluxes and wavelengths for filters with multiple measurements.

Also shown in the figure is
a theoretical spectrum from the Spanish Virtual Observatory\footnote{http://svo2.cab.inta-csic.es/theory/main/} (SVO) and  synthetic photometry from the folding of this spectrum through the observed bandpasses. No extinction correction is applied, as it is likely quite small (see
Table~\ref{tab:TMR} below; $A_{\rm V}$ is estimated to be less than $0.1$).
The theoretical spectrum is based on the ATLAS9 stellar atmospheric models of \cite{castelli03}, which contain synthetic spectra for a three-dimensional grid of effective temperatures, gravities and metallicities. We adopted a grid point that is closest to the stellar parameters derived by LAMOST ---effective temperature $T_{\rm eff}=6250$\,K, $\log g=4$, and metallicity $-0.5$. 
Based on the parallax of $1.685$ mas from {\it Gaia} \citep{gaia16, browngaia2021}, we adopt a stellar radius  $R_*=1.9 \frac{d}{600 pc} R_{\odot}$, a value that yields the best fit to the observed broad-band photometry. 

Evidently, the theoretical spectrum agrees well with the weighted-averaged photometry.
%
%
%
We note, however, that one cannot rule out a contribution to the UV flux from a single WD companion. The flux from
even a very hot ($T_{\rm eff}=10^5$ K) WD with the smallest mass consistent with our companion-mass constraints (and hence the largest allowed WD radius; $M =1.1 M_{\odot}$, $\log g = 9$) could still be accommodated within the observed broad-band photometry at the $\sim 10$\% level.
Future FUV observations, at wavelengths where the F-star contribution
is negligible, can better rule out hot, massive WD companions.

\subsection{Stellar parameters of J1255}
\label{sec:J1255_MRT}


To estimate the mass, radius, metallicity and age of the primary star, we used the  python \texttt{ISOCHRONES}\footnote{https://isochrones.readthedocs.io/en/latest/} software package \citep{morton15}, used by many studies
\citep[e.g.,][]{huber16,mathur17,koposov20}. 
The software uses the MESA Isochrones and Stellar Tracks
(MIST) models \citep{choi16} to determine the stellar  physical parameters, based on the observed photometric and spectroscopic parameters. We used the available photometry from the UV to the IR bands, with a prior on temperature and metallicity from the LAMOST spectrum analysis. The SPC analysis  \citep{SPC_21,buchhave12} of the TRES spectra, yielded similar results, except for the stellar metallicity, as seen in Table~\ref{tab:atmospheric}.  
The photometric data were combined with four additional constraints: a parallax of $1.685 \pm 0.012$ mas 
(a distance of $595\pm 5$ pc), 
as reported in the \textit{Gaia} EDR3 archival database,\footnote{https://gea.esac.esa.int/archive/} and the spectroscopic parameters that we used above in Section~\ref{sec:J1255_SED}:
$T_{\rm eff} = 6250 \pm 100$ K, 
$\rm \log g = 4.0 \pm 0.1$
and $\rm [Fe/H] = -0.50 \pm 0.05$. 

\begin{figure*} 
\centering
\includegraphics[scale=0.5]
{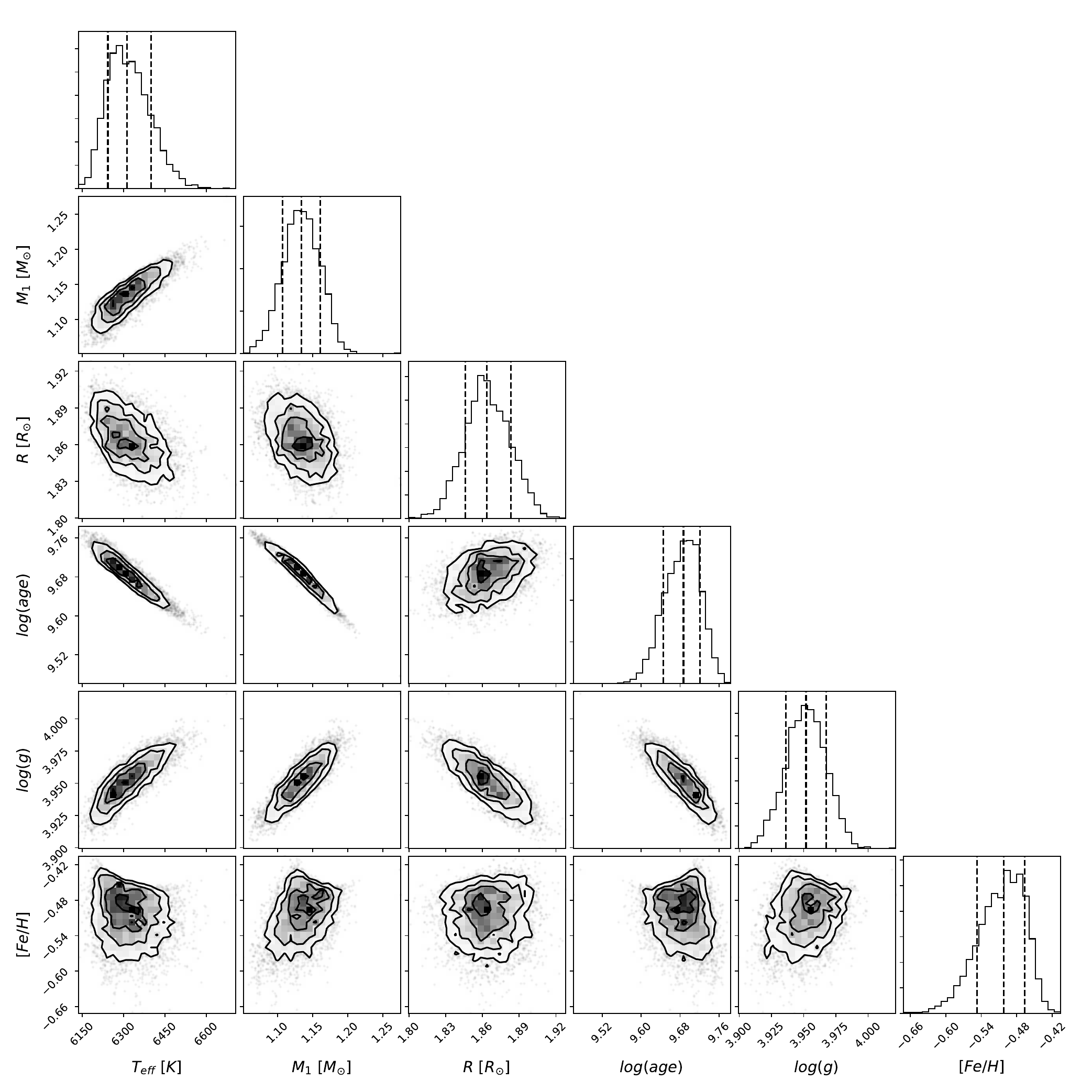}
\caption{ 
MCMC posterior distributions
for the  J1255 stellar physical parameters: $T_{\rm eff}$ [K], $M_1 [M_{\odot}]$, $R \ [R_{\odot}]$, log of stellar age, $\log \rm g$ and $\rm [Fe/H]$.  See Fig.~\ref{fig:corner} for details.
}
\label{fig:mcmcPhys}
\end{figure*}

\begin{figure}
\centering
\includegraphics[scale=0.6]
{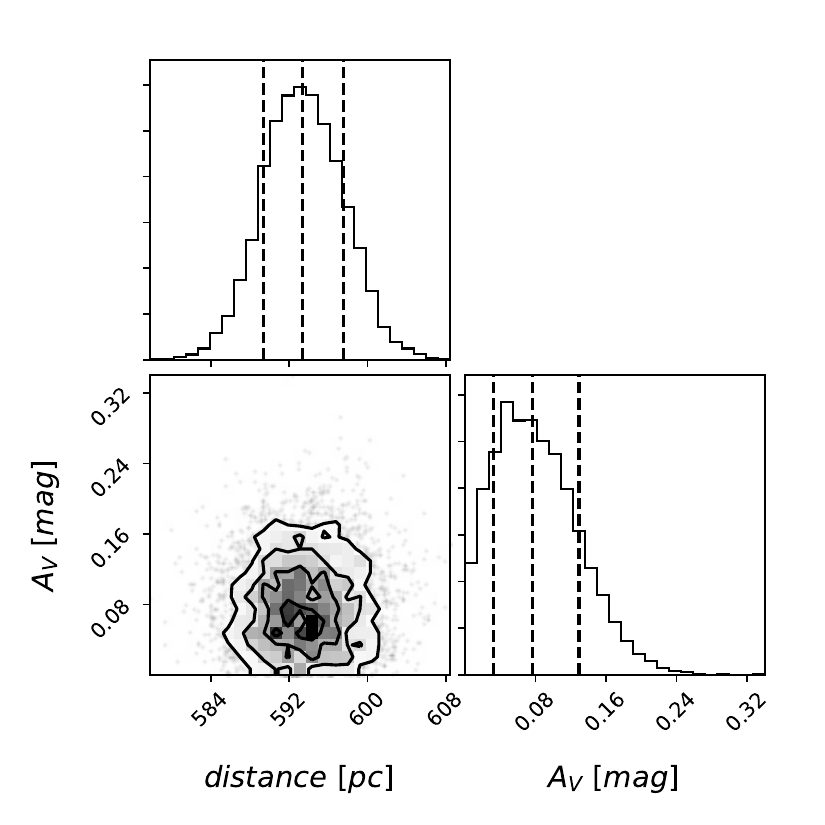}
\caption{ 
MCMC posterior distributions
of J1255 distance and V-band extinction, $A_V$. See Fig.~\ref{fig:corner} for details.
}
\label{fig:mcmcExt}
\end{figure}

The code explores a five-dimensional parameter space of initial mass, stellar age, metallicity, distance and line-of-sight V-band extinction, using an MCMC algorithm, given the priors listed in Table~\ref{tab:TMR}. The resulting corner plots for the physical parameters are shown in Figs.~\ref{fig:mcmcPhys}--\ref{fig:mcmcExt}.
We note that unlike many X-ray binaries, for which the optical star fills its Roche lobe \citep[e.g.,][]{orosz07, orosz09}, the stellar ellipsoidal shape of J1255 cannot affect our results, as the stellar distortion is minute. The ellipsoidal modulation is less than $0.5$\%, as a result of the small Roche-lobe filling factor of 0.43 (see Section~\ref{sec:Mass} below). 
%

The adopted parameters of the primary of J1255 are listed in Table~\ref{tab:TMR}.  We have conservatively assumed that the uncertainties reported by \texttt{ISOCHRONES} are slightly too small, because it ignored some systematic errors,
in particular those of the MESA Isochrones and MIST models.
We assumed $\sim5\%$ uncertainty in the primary mass, as the \texttt{ISOCHRONES} estimation depends on stellar models, and $0.05R_{\odot}$ in the stellar radius. The latter is directly constrained by the broad-band photometry and the {\it Gaia} distance.

To study the Galactic dynamics of J1255, we have derived its 
Galactic orbit in Appendix~\ref{sec:galactic} and Fig.~\ref{fig:galacticOrbit}.
The star moves on a constrained Galactic radial orbit between 
$7.8$--$10.5~\mathrm{kpc}$ that takes it up and down the plane by  $Z\sim0.6~\mathrm{kpc}$. Using estimates of the Galactic velocities, we further find
$(U,V,W) = (-35.08\pm0.26, 5.99\pm0.14, -17.44\pm0.28)$ km~s$^{-1}$,
suggesting that J1255 
has typical thin-disk kinematics \citep[e.g.,][]{bashizucker2022}.

\begin{table}
\caption{Adopted Stellar Parameters of the primary of J1255
\label{tab:TMR}}
\begin{center}
\begin{tabular}{l l l  l | l}
\hline
Parameter & Units & Prior & \texttt{ISOCHRONES}  & Adopted\\
\hline
$T_{\rm eff}$ & K   &  $\mathcal{N}(6250,100)$   &  $6310^{+90}_{-70}$  
&  $6300 \pm 100 $   \\
M  & $M_{\odot}$  & $\mathcal{U}(1.0,1.6)$ & $1.13^{+0.03}_{-0.02}$
& $1.13\pm 0.05$ \\
R  & $R_{\odot}$  & $\mathcal{U}(0,100)$  & $1.86^{+0.02}_{-0.01}$ 
& $1.86 \pm 0.05$ \\
$\log( \rm {age})$  & yrs & $\mathcal{U}(9,10)$ & $9.69^{+0.03}_{-0.04}$ 
& $9.70\pm 0.05$ \\
$\log {\rm g}$  &  c.g.s.  & $\mathcal{N}(4.0,0.1)$ &  $3.95^{+0.02}_{-0.01}$ 
& $3.95 \pm 0.05$ \\
$\rm [ Fe/H ]$     &  &   $\mathcal{U}(-0.75,-0.25)$     & $-0.50^{+0.03}_{-0.05}$ 
 & $-0.50 \pm 0.05$\\
Distance       & pc  &  $\mathcal{U}(580.8, 606.0)$    &  $593^{+5}_{-4}$  &  $593\pm 10$\\
$A_V$          &       & $\mathcal{U}(0, 1)$  &  $0.077^{+0.052}_{-0.045}$    
& $0.08\pm 0.05$ \\
\hline
\end{tabular}
\end{center}
\end{table}

\section{The Mass of the Unseen Companion}
\label{sec:Mass}

To derive the mass of the unseen companion, one can use the RV amplitude, which depends, for a circular orbit, on the primary and secondary masses, $M_1$ and $M_2$, the period, $P$, and the inclination $i$:
%
\begin{equation}
  \label{eq:K}
  K            =213 \, \,
  \bigg(\frac{M_1}{M_\odot}\bigg)^{1/3}
  \bigg(\frac{P}{\text{day}}\bigg)^{-1/3} 
  \frac{q}{(1+q)^{2/3}} \  \sin i \quad \text{km~s$^{-1}$} \ ,  
\end{equation}
%
where $q\equiv M_2/M_1$. 
We already derived an estimate for $M_1$, so we are left with only one unknown --- the orbital inclination, in order to be able to solve for $M_2$. 

To derive the inclination one can assume the stellar rotation is aligned and synchronized with the orbital motion, as
might be expected due to long-term tidal interaction between the close components \citep[e.g.,][]{mazeh08}. We then compare the observed ellipsoidal-modulation amplitude with theoretical expectations for such a short-period binary, yielding one constraint on the companion mass and orbital inclination.
Another constraint can come from the projected equatorial rotational velocity of the F star, which can be determined from the observed spectral-line broadening,
provided we know the stellar radius. 
The advantage of using both constraints, the projected stellar rotation and the ellipsoidal amplitude, is the different dependence of the two on the inclination ---- while the projected rotational velocity depends on $\sin i$, the ellipsoidal amplitude depends on $\sin^2i$ \citep[e.g.,][]{morris93,faigler11}.

In the case of J1255, the  {\it TESS} photometry (Section~\ref{sec:TESS}) shows a flux modulation of $4.27 \pm 0.05$\,ppt,  
with a period that is exactly half the RV period, and with a phase that is 
expected for the ellipsoidal
modulation, with no indication of an eclipse.  Analysis of the TRES and CARMENES spectra yields an estimation for the projected rotational broadening of 
$32.3\pm0.5$ km~s$^{-1}$.  
Neither of these two constraints is iron-clad. 
The expected ellipsoidal modulation is based on theoretical approximations that are known to fail in reproducing the fine details of the effect \citep{gomel21a}, and the determination of the projected equatorial rotation velocity requires assumptions about how to account for line-broadening effects
other than rotation \citep[e.g.,][]{barklem00,atmospheric16}, and further assumptions about synchronization and alignment of the stellar rotation with the orbital motion. 
Nevertheless, we have used these two additional constraints with enlarged uncertainties of 10\% that reflect these associated open questions.


To obtain the secondary mass and inclination of J1255, we use the
{\tt PyMC} package for probabilistic programming \citep{salvatier2016}. The search 
uses uninformative priors on the cosine of the inclination, $\cos i$, and the mass ratio $q$  and the orbital period $P$, as listed in the top part of 
Table~\ref{tab:getCompanionMass}.
Given these priors, we calculate the values of  secondary mass and inclination by comparing the resulting projected rotational velocity, ellipsoidal amplitude and RV semi-amplitude, as found from Eq.~\ref{eq:K} above, together with 
%
\begin{equation}
  \label{eq:vsin1}
  {\rm v}\sin i =50.6 \,  
  \bigg(\frac{R_1}{R_\odot}\bigg) 
  \bigg(\frac{P}{\text{day}}\bigg)^{-1} 
  \sin i \quad \text{km~s$^{-1}$},
\end{equation}
\begin{equation}
  \label{eq:ell}
  A_2        = 13.44\ \alpha \,
  \bigg(\frac{M_1}{M_\odot}\bigg)^{-1}
  \bigg(\frac{R_1}{R_\odot}\bigg)^{3}
  \bigg(\frac{P}{\text{day}}\bigg)^{-2} 
  \frac{q}{1+q} \,\,  \sin^2i\quad \text{ppt} \, ,  \\
\end{equation}
%
where $\alpha$ is the ellipsoidal coefficient \citep{faigler11,gomel21}, which we set to $1.3$, given the temperature of the primary. 
We then calculated the likelihood of the corresponding observed quantities, assuming the errors are Gaussian and additive, as listed in the lower part of Table~\ref{tab:getCompanionMass}.
We assume the inclinations of the orbit and the stellar rotational axis are both $i$, and the rotation period and the orbital period both equal $P$.
 
Table~\ref{tab:getCompanionMass} lists the resulting posterior median and $1\sigma$ widths of the system parameters which, based on the adopted uncertainties, are consistent with the input distributions. 
This indicates the validity of our assumptions, given the problem is over-determined, as we have in hand more constraints than free parameters.
In particular, the primary radius turns out to be only $0.166$ of the binary separation, and the star fills only $0.43$ of its Roche lobe. In such cases, Equation~\ref{eq:ell} is a very good approximation of the ellipsoidal modulation.

%

\begin{table*}
\caption{Parameters (priors) used to derive the unseen companion mass and their posterior distributions}
\label{tab:getCompanionMass}
\begin{center}
\begin{tabular}{l l l l | l}
\hline
Parameter & &Units & Prior & Posterior \\
\hline
 Primary mass & $M_{1}$ &  $M_{\odot}$   & $\mathcal{N}(1.13,0.05)$ & $1.16\pm0.05$\\
Primary radius & $R_{1}$ &  $R_{\odot}$   &  $\mathcal{N}(1.86,0.05)$& $1.82\pm0.05$     \\
Orbital period & $P$ &  day   &  $\mathcal{N}(2.760549,0.000013)$  & $2.760549\pm0.000013$   \\
Inclination angle & $\cos{i}$ &     &  $\mathcal{U}(0,1)$  & $0.54^{+0.10}_{-0.16}$   \\
Mass ratio & $q$ &     &  $\mathcal{U}(0.5,2)$   & $1.2^{+0.2}_{-0.17}$  \\
\hline
Projected rotational velocity& ${\rm v}\,{\sin i}$& km~s$^{-1}$ &$32.3\pm 3.0$& $28.3\pm2.2$ \\
Ellipsoidal amplitude & $A_2$& ppt &$4.27 \pm0.43$ & $4.71\pm 0.36$\\
RV semi-amplitude & $K$& km/s & $96.1 \pm 0.2$& $96.1 \pm 0.2$\\
\hline
\end{tabular}
\end{center}
\end{table*}

\begin{figure}
\centering
\includegraphics[scale=0.4]{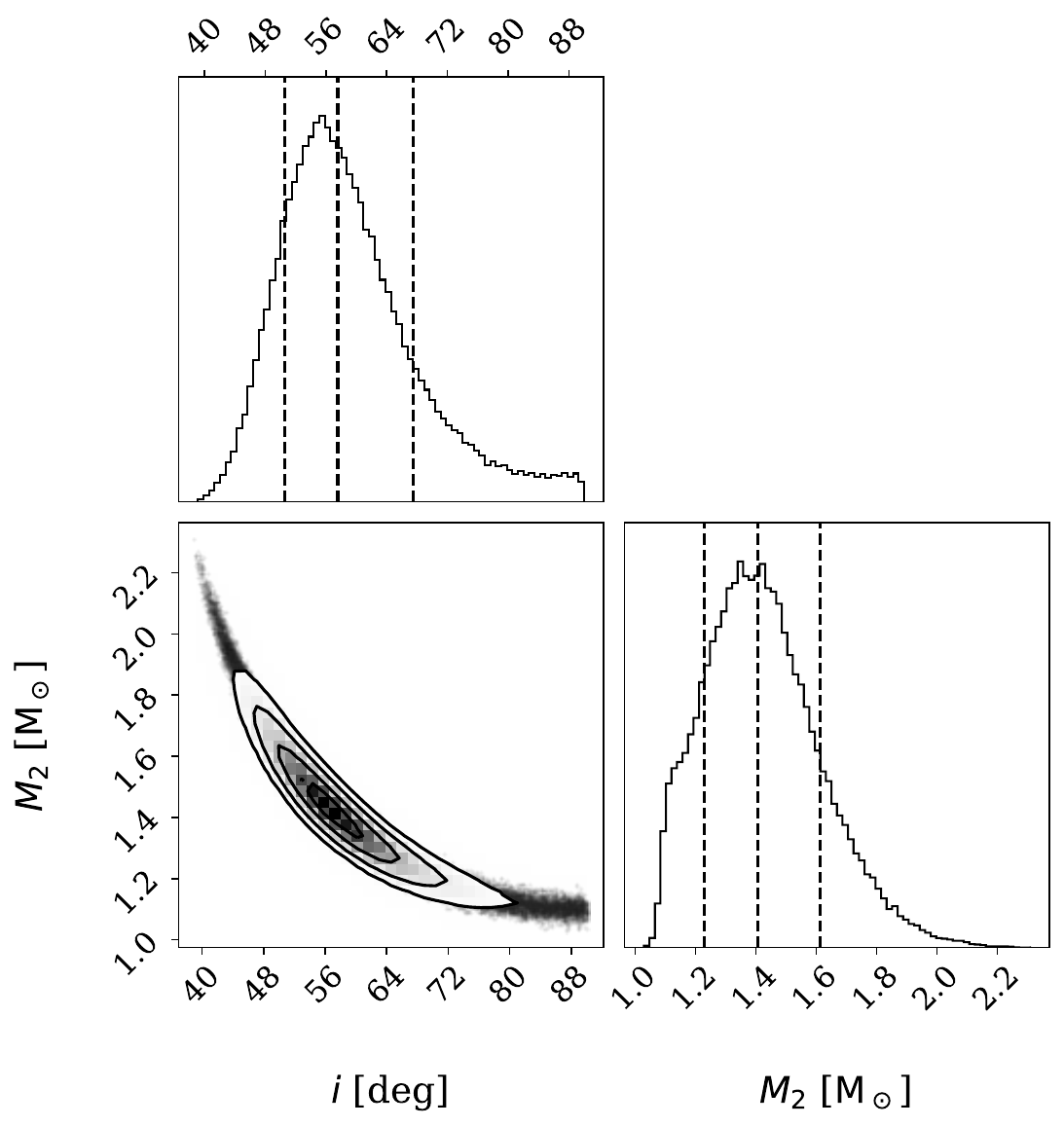}
\caption{Possible values of the secondary mass vs.~the orbital inclination, resulting from the MCMC run. 
The off-diagonal plot presents the joint distributions with contours, and the diagonal elements show the probability distribution of the two parameters, with percentiles of $16\%$, $50\%$ and $84\%$, denoted by vertical dashed lines.
}
\label{fig:M2_sini}
\end{figure}

Fig.~\ref{fig:M2_sini} shows the results of
the MCMC process for the companion mass
and its orbital inclination.
From the 1-d marginalized posterior, the mass estimate for the unseen companion is
\begin{equation}
1.41^{+0.21}_{-0.18}\, M_{\odot}\,.
\end{equation}
We note that the posterior mass distribution is highly asymmetric, with a sharp drop at $\sim 1.1 M_{\odot}$\,, and a long tail that extends up to  $\sim 2 M_{\odot}$. The $3\sigma$ confidence-level mass range is

\centerline{  $1.07$ -- $2.08\, M_{\odot}\,$.}

\section{Discussion}
\label{sec:discussion}

We have shown that the unseen companion of J1255 is more massive than the visible F-type star, with a mass of 
$1.1$--$2.1\, M_{\odot}$ at the $3\sigma$ confidence level, and a mass-distribution median of $1.4\, M_{\odot}$. 

This mass estimate is based on the observed RV modulation, the amplitude of the ellipsoidal modulation, and the rotational broadening of the spectral lines.
As emphasized above in Section~\ref{sec:Mass}, the derivation relies on several assumptions, particularly
the alignment and synchronization of the 
primary with the orbital motion, as expected for such a short-period binary \citep[e.g.,][]{mazeh08}, even more for an {\it evolving} sub-giant. 
We further note that the expression used for the expected ellipsoidal modulation is based on theoretical approximations, which is inherently inaccurate to a certain degree \citep{gomel21a}.
On the other hand, using the ellipsoidal modulation and the rotational broadening constraints makes the problem {\it over-determined}. The mere existence of  a solution for the secondary mass and the inclination is strong evidence that our assumptions are applicable to J1255.

As shown, the synchronization assumption, implying that the F star is rotating around its axis with a period of 2.76 days, can also account for the observed X-ray flux, $\sim 1.5\times 10^{-14}$ mW/m$^2$ \citep{4XMM_20}, measured from the XMM-Newton serendipitous observations, as such a flux at the stellar distance is typical for F stars with this rotational period. 
Another set of serendipitous observations was performed with the {\it Swift}\footnote{https://swift.gsfc.nasa.gov/}
UVOT telescope \citep{swift04} 
in the UVM2 band, revealing a modulation with the period and phase of the {\it TESS} modulation, with an amplitude of $8.4\pm2.1$\,ppt. 
%
The modulation reported here  is likely the first detection of an ellipsoidal modulation in the UV. 

%
The stellar UV brightness is coming from the upper layers of the stellar atmosphere, which are tidally distorted by the unseen companion. 
  
\subsection{Nature of the more-massive secondary in the J1255 system}

Could the more-massive companion in J1255 be another MS star? In such a case we would have seen a second set of spectral lines with the same, or even higher, signal-to-noise ratio than that of the F star. 
This is especially true for the spectra obtained near quadrature phases, when the two sets
of lines would have been well separated, as the RV difference between the two objects is $\sim 200$ km~s$^{-1}$. 
No trace of a second star is seen in the TRES and CARMENES spectra, 
despite their high resolution and high SNR, turning  the system to be 
a single-lined spectroscopic binary (SB1). 

SB1 with a mass ratio larger than unity most likely hosts a dormant (i.e. detached non-interacting) compact-object companion, as 
in a coeval binary with two MS stars, we would expect the more massive star to outshine its companion. 
The fact that we observe the less massive star in the system suggests that the other star is a compact object. 


However, this is not the case for all SB1s. Algol-type binaries, which probably have gone through a mass-transfer phase during their evolution \citep{erdem14,dervi18,chen20}, are famous 
counterexamples. In these binaries, a giant or a sub-giant is the less-massive component, and yet is the brighter star in the system \citep{nelson01, budding04, mennekens17, negu18}. Therefore, SB1s with a photometric primary in advanced stages of its evolution probably have a more-massive stellar companion as a photometric secondary. Algol-type binaries are much more common than binaries with a neutron-star or a black-hole component.

Indeed, a number of systems recently proposed to consist of an evolved primary and a dormant compact-object secondary \citep[e.g.,][]{thompson19, liu19, rivinius20, jayasinghe21} are more likely to be binaries with an evolved primary and an MS secondary  \citep[see discussion by][]{van-den-Heuvel20, irrgang20, shenar20, bodensteiner20, mazeh20, el-badry20, el-badry22a, el-badry22b}. The works of \cite{BH18,BH19}, \cite{saracino22} and \cite{El-Badry22} stand in contrast to most of these recent suggestions, with their primary on the MS, and therefore might have a genuine unseen black-hole companion. J1255 presents us with a similar situation, as demonstrated above, where the secondary cannot be an MS star. 
In the case of J1255, as the visible star is only slightly evolved, it cannot hide in its glare a MS star with a mass larger than the primary, and therefore the secondary is probably a compact object. 

\subsection{J1255 secondary --- a probable neutron star}

The mass we have estimated for the unseen companion in J1255 is within the range of measured masses of confirmed neutron stars

\citep[e.g.,][]{lattimer2012,lattimer14,
ozel16}.\footnote{https://stellarcollapse.org/nsmasses.html}
%
%
We therefore argue that the evidence presented here strongly points to a neutron-star companion in J1225.


Could the unseen companion be some other type of compact object? Although the companion is not massive enough to be the BH remnant of a massive star \citep[e.g.,][]{BlackCat16}, it could be an unusually massive single white dwarf.  A recent analysis of the Gaia EDR3 catalog \citep{GentileFusillo21} confirms the long-known picture that the mass distribution of white dwarfs in the solar neighborhood has a long tail extending to $1.2 M_{\odot}$ and even higher. 
The range of allowed  companion masses in J1255 thus includes the possibility of such a massive white dwarf.
Similar ambiguities regarding the identification of a {\it dormant} neutron star in short-period binaries are bound to affect future similar discoveries, unless deep UV observations can rule out the white-dwarf alternative.

\subsection{An inner close binary --- an alternative model?}
\label{sec: triple}

Could the unseen companion in J1225 be, in itself, a close binary with a period of, say, $0.2$ d, composed of two low-mass stars, each with a mass of $\gtrsim 0.6 M_{\odot}$?
The five CARMENES spectra, which include the red part of the spectrum, allow us to explore this possibility.  Applying a 
\texttt {TODCOR} analysis \citep{TODCOR94}, which is tuned to find secondary lines in a spectrum even when they are faint, we identify no trace of a set of lines typical of M stars, down to a level of 0.5\% of the intensity of the F star. This upper limit rules out any M star with a mass exceeding $\sim 0.5 M_{\odot}$ (See Appendix~\ref{sec:CARMENES}).
Still, one could hypothesize that J1255 is a hierarchical triple system with a close-binary secondary, composed of two or at least one white-dwarf component. For example, two white dwarfs, each with a typical mass of $0.7\,M_{\odot}$, or a massive white dwarf with a mass of, say,  $1 \,M_{\odot}$, and an M star with $0.4\,M_{\odot}$, are two possibilities that the CARMENES spectra cannot rule out, due to the low-luminosity of the white dwarf and/or the  low-mass M star.
%

Nevertheless, we 
argue that the scenario in which J1255 is a triple system, where the F star orbits a close binary with a white dwarf as one of its components, is unlikely.
Let us first consider the dynamical stability of a hierarchical triple system. 
The stability depends primarily on $\mathcal{R}=P_3/P_{1,2}$\,, where  $P_{1,2}$ is the orbital period of the close binary and  $P_3$ is that of the wider orbit. Other factors are the mass ratios of the close and wide pairs, the mutual inclination and the eccentricities of the two orbits.  
Many studies of hierarchical-triple stability suggest that for a system to be stable $\mathcal{R}$ has to be at least $5$--$10$ \citep[e.g.,][]{orlov00,mardling01}. In fact, the known triple systems with short $P_2\lesssim 0.25$\,d all have $\mathcal{R}\gtrsim 100$ \citep{tokovinin18}. Triple systems having $\mathcal{R}$ that is too small go through a dynamical interaction that eventually destroys the system \citep{toonen21}. This is probably why the most compact known triple systems are with $P_3$ of $\sim 30$\,d \citep{tokovinin18, borkovits22}. 

Even if we assume that J1255 is a triple system with an extremely short $P_3$  of $2.76$\,d, the stability criterion implies for the presumed close binary of J1255 that $P_{1,2} \lesssim  0.25$ d, or a separation smaller than $\sim 2\, R_{\odot}$\,, with a Roche-lobe radius of each of the components on the order of $1\,R_{\odot}$\,. 
To allow some space for the presumed M star, the hypothesized close binary could not be much more compact, unless it is composed of two white dwarfs. 
A system with  $\mathcal{R}\sim 5$ is on the edge of being stable, and any evolutionary process that slightly decreases $\mathcal{R}$ can render the system dynamically unstable.   

In any event, a close binary with a separation of $\sim 2\, R_{\odot}$ does not have enough space for a star to evolve through the giant phase and form a white dwarf. Therefore, the original separation of the close binary must have been substantially larger, so that the close binary shrank to its present dimension either by mass transfer or a common envelope phase. 
Usually, it is expected that in the adiabatic mass-transfer evolution phase of a triple system, the inner orbit shrinks while the outer orbit expands \citep{toonen21},  leading to an increase of $\mathcal{R}$. 
This means that $\mathcal{R}$ was originally even smaller than its present value. Such a system could not have survived the dynamical interaction between the two orbits.

Another possible route of evolution is that the three stars were engulfed by the expanding atmosphere of the presumed progenitor of the white dwarf and shrank to the present configuration. However, as shown by \citet{comerford20}, who considered the outcome of a triple system in a common-envelope phase, such a system is doomed to be destroyed by the chaotic motion inside the extended envelope. 

In summary, it is difficult to imagine an evolutionary path for J1255 that forms such a close hierarchical triple with a white-dwarf component. A scenario that led to a double white-dwarf close binary, with an F star as the tertiary star at a period of $2.76$\,d is even less probable. 
We, therefore, conclude that the hierarchical triple-system assumption for J1255, with the secondary being a close binary, is quite unlikely, because of the short orbital period of the F star. 

\subsection{A wide companion in the system?}
Another triple-system possibility we have considered is that the $2.76$\,d binary is orbited by a faint distant companion. Such a wide companion will not affect the neutron-star conclusion for the companion, but it can put the J1255 binary evolution in a different context. The third star could have been an angular-momentum sink for the short binary, and pushed the J1255 system by tidal friction into a relatively short period \citep[e.g.,][]{MazehShaham79,fabrycky07,naoz16}. A triple system model for a binary with a neutron star was first suggested by \citet{MazehShaham77} for the X-ray binary Her X-1.  Although the idea for Her-X1 is probably wrong \citep[e.g.,][]{truemper86,lanzafame94}, the  triple-system model is applicable to several close binaries with a compact companion
\citep[e.g.,][]{bailyn87,Triple07}.

We have performed two searches for a hint of a wide companion to J1255, both with negative results. First, as explained in Section~\ref{sec:orbital}, we looked for a long-term trend in the RV residuals, with upper limits of $-0.0015 < a < 0.0030$ km~s$^{-1}$/day at the $2\sigma$ level. This rules out a distant M-star companion of $\sim0.5\,M_{\odot}$ up to a distance on the order of 10 au, unless we caught the wide binary in a special orbital phase.  

Second, we performed 
infrared high-resolution adaptive optics (AO) imaging of J1255 at Palomar Observatory with the PHARO instrument \citep{hayward2001}, using the natural guide star AO system P3K \citep{dekany2013}, as detailed in Appendix~\ref{sec:AO}.  The range of sensitivity to companions is only on the order of >1 arc-sec, equivalent to 600 au at the distance of J1255. This is much larger than the scale expected for wide binaries that could modify the evolution of the close binary by dynamical friction. Nevertheless, a third star at this distance would suggest a more complicated hierarchical system.
However, we did not detect any wide companion above the observational threshold detailed in Appendix~\ref{sec:AO}.

The two negative results do not cover the whole range of possible distances of a wide companion, but, nevertheless, are consistent with the postulation that the J1255 binary system does not have a third distant companion.


\subsection{The J1255 binary in context}

The J1255 system probably consists of an F-type star in orbit with a dormant neutron star, which was in the past a young pulsar. 
Currently, the optical star is far from filling its Roche lobe, with a filling factor smaller than 0.5 (see Section~\ref{sec:Mass}).
In the future, when the F star further evolves and transfers mass onto the compact companion, J1255 might turn into a bright X-ray source, and could 
later
even host a millisecond pulsar.

The discovery of this interesting system, devoid of excess X-ray emission, was made possible by the large RV survey performed by the LAMOST multi-object spectrograph, followed by accurate photometry and high-resolution spectroscopy. The discovery relies heavily on the fact that the observed star is only a slightly evolved F star, and not a giant or a supergiant, as in the case of Algol-type binaries \citep{budding04}. 
In systems with evolved stars \citep{nelson01}, the bright star can easily hide a MS or sub-giant companion in its glare \citep[e.g.,][]{NGC_1850_21, NGC_1850_NO_21,el-badry22a,el-badry22b}.

We do expect many dormant systems with either a neutron-star or a black-hole companion, as the dormant phases of such binaries are much longer than the young-pulsar and X-ray accretion phases. In our system, for example, the age of the F star, as derived by \texttt{isochrones}, is about $5$ Gyr, while the pulsar phase, during which the object was spinning as a pulsar, is estimated to last on the order of $10$--$100$ million years \citep[e.g.,][]{PhinneyBlandford81}, and the mass transferring phase of the F %
star is substantially less than 1 Gyr.
To find additional dormant systems one needs similar large RV surveys, as these close binaries are quite rare. 
Based on our {\it single} finding, a preliminary guess for the frequency of the dormant short-period systems is on the order of one to a few per million stars. 

To find more systems with compact objects one can use the {\it Gaia} large sample of spectroscopic binaries  \citep{NSS} released very recently, which contains $181\,327$ orbits, including the orbit of J1255. This sample probably has other systems with compact components, as suggested by the {\it Gaia} binary paper \citep[but see][for a careful assessment of some of these systems]{el-badry_NSS22}.
%
%
Another approach is to search for MS stars that exhibit a relatively large ellipsoidal modulation, which, again, might indicate a massive unseen companion \citep{gomel21c}. Photometric light curves are available for a large number of stars \citep{rowan21}, on the order of $10^{8}$, or even more.

The spectroscopic and the photometric approaches alike necessitate RV follow-up observations to establish the mass of the companion and its nature.
With the results of such surveys one will be able to study the frequency and other statistical features of the dormant compact objects, neutron stars and  black holes, residing in {\it close} binary systems. Both populations are important for understanding the evolution of binaries with massive companions \citep[e.g.,][]{belczynski02} that can lead to the generation of neutron stars or black holes, and their possible contribution, in extreme cases, to the merger events detected recently via gravitational waves. 

\section*{Acknowledgements}

%
%
We thank the anonymous referee for thoughtful comments and suggestions that improved the original manuscript, and H-W Rix for discussion of the beaming modulation.
This research was supported by Grant No. 2016069 of the United States-Israel Binational Science Foundation (BSF) to TM, 
Grant No. I-1498-303.7/2019 of the German-Israeli Foundation for Scientific Research and Development (GIF) to TM and DM,  
a grant from the European Research
Council (ERC) under the European Union’s FP7 Programme, Grant No.
833031 to DM,
the National Key R\&D Program of China (No. 2019YFA0405100) to SD. The research of NH and SS is supported by a Benoziyo prize postdoctoral fellowship.

This paper uses data from the LAMOST survey: Guoshoujing Telescope (the Large Sky Area Multi-Object Fiber Spectroscopic Telescope, LAMOST) is a National Major Scientific Project built by the Chinese Academy of Sciences. Funding for the project has been provided by the National Development and Reform Commission. LAMOST is operated and managed by the National Astronomical Observatories, Chinese Academy of Sciences.

This work has also made use of data from the European Space Agency (ESA) mission \textit{Gaia} (https://www.cosmos.esa.int/
gaia), processed by the \textit{Gaia} Data Processing and Analysis Consortium (DPAC; https://www.cosmos.esa.int/web/gaia/
dpac/consortium). Funding for DPAC has been provided by national institutions, in particular the institutions participating in the \textit{Gaia} Multilateral Agreement.

This research made use of \texttt{exoplanet} \citep{foreman21} and its dependencies \citep{astropy13,astropy18, salvatier2016, team2016theano, luger2019starry, agol2020analytic}. 

\section*{Data Availability}

Data used in this study are available upon request from the corresponding
author.



\bibliographystyle{mnras}
\bibliography{MNRASbib} 



\appendix


\section{J1255 Galactic dynamics}
\label{sec:galactic}

To study the Galactic dynamics of J1255 we derived its current Galactic position, using its equatorial coordinates, parallax, and proper motions as given by {\it Gaia} EDR3 \citep{browngaia2021}, and our derived center of mass RV. We used the \texttt{GALPY} package\footnote{https://docs.galpy.org/en/v1.7.1/} for Galactic motion calculations \citep{bovy2015} with the \texttt{MWPotential2014} approximation for the Milky Way gravitational field \citep{bovyrix2013} to back-propagate the stellar orbit $4.8~\mathrm{Gyr}$, using our best estimation of the stellar age. 

\begin{figure} 
\centering
\includegraphics[scale=0.45]
{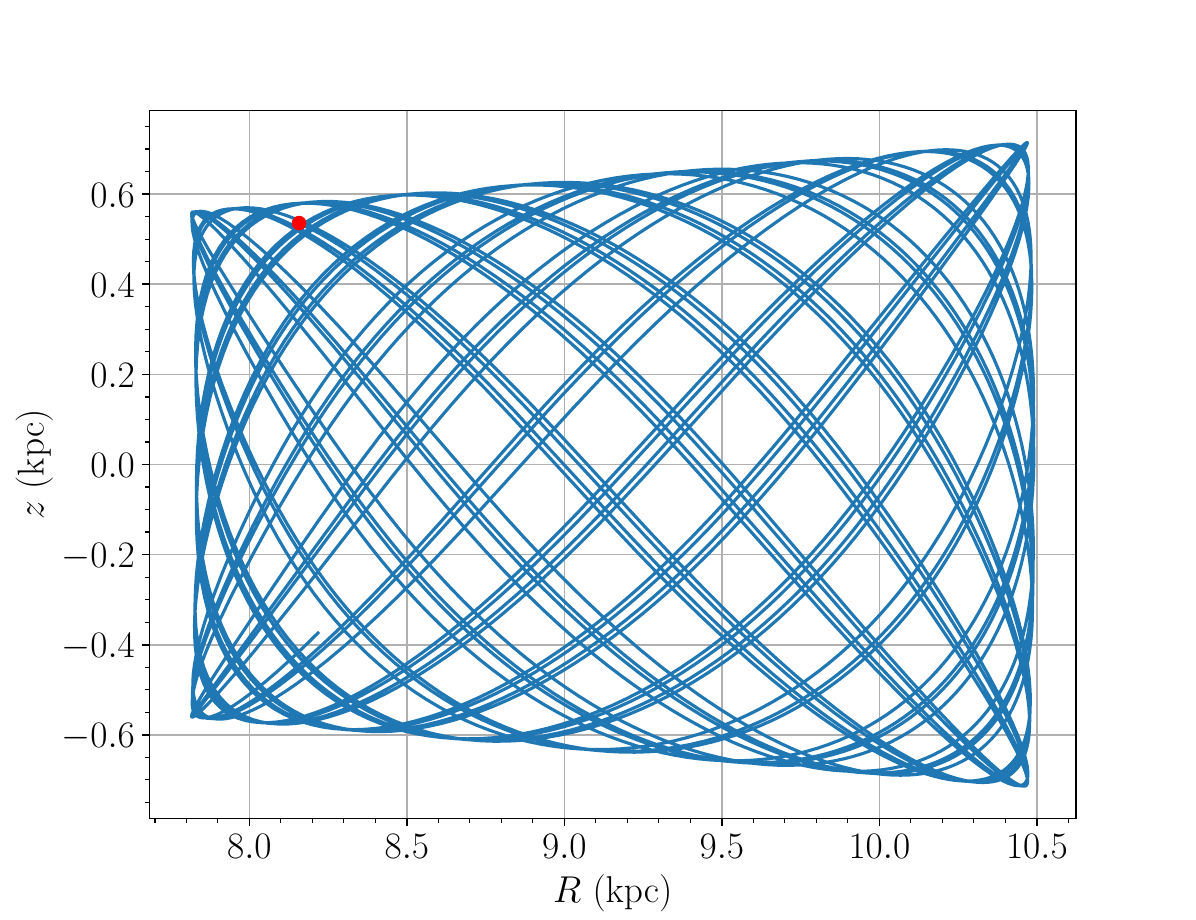}
\caption{J1255 stellar Galactic scale height above mid plane $Z$ as function of Galactic radii with respect to Galactic center. Blue curve marks the orbit from current position (red point) and back in time for $4.8$ Gyr.}
\label{fig:galacticOrbit}
\end{figure}

The resulting Galactic orbit is plotted in 
Fig.~\ref{fig:galacticOrbit}. The star moves on a constrained Galactic radial orbit between 
$R=7.8$--$10.5~\mathrm{kpc}$ that takes it up and down the plane to a scale height of $Z\sim0.6~\mathrm{kpc}$, well inside the galactic disk. 

\section{Searching for a faint companion of J1255}

\subsection{Searching for a faint spectroscopic component in the CARMENES spectra}
\label{sec:CARMENES}

Search for a secondary component in the CARMENES spectra was performed with \texttt{TODMOR} \citep{Zucker2003,Zucker2004}, an implementation for multi-order spectra of the two-dimensional cross-correlation method \texttt{TODCOR} \citep{Zucker1994}. This technique uses two different templates, one for each component, scaled by their
flux ratio, to construct a two-dimensional cross correlation as a function of the RVs of the two templates. The peak of the CCF yields the RVs of the two components of the binary. 

\texttt{TODMOR} can be used to exclude a faint secondary in the spectrum, as was done here. We inspected the CARMENES spectra with the highest SNR to search for spectral signatures of any companion, using an assortment of templates with a range of $6000$--$4000$ K,  obtaining negative results. This included the whole spectra, as well as individual echelle orders, all with the same negative results.

\subsubsection{Injection-retrieval tests}

To quantify the negative result, we performed an injection-retrieval test to determine the limiting flux ratio at which a secondary star could have been detected. We used the highest SNR spectrum from CARMENES, and injected a PHOENIX synthetic spectra of 4000\,K, Doppler-shifted at $-200$ km~s$^{-1}$. The injected spectra were scaled using different flux ratios from 0.5\% to 5\%, and broadened with the expected rotational projected velocity of a synchronized K7 star, of $10$\,km~s$^{-1}$. 

We inspected the resulting spectra with \texttt{TODMOR} to see if we could detect the injected signal, using a primary template with the same parameters as the ones used to compute the RVs, and a secondary template with the parameters used in the injected spectrum.

\begin{figure}
\centering
\includegraphics[width=0.45\textwidth]{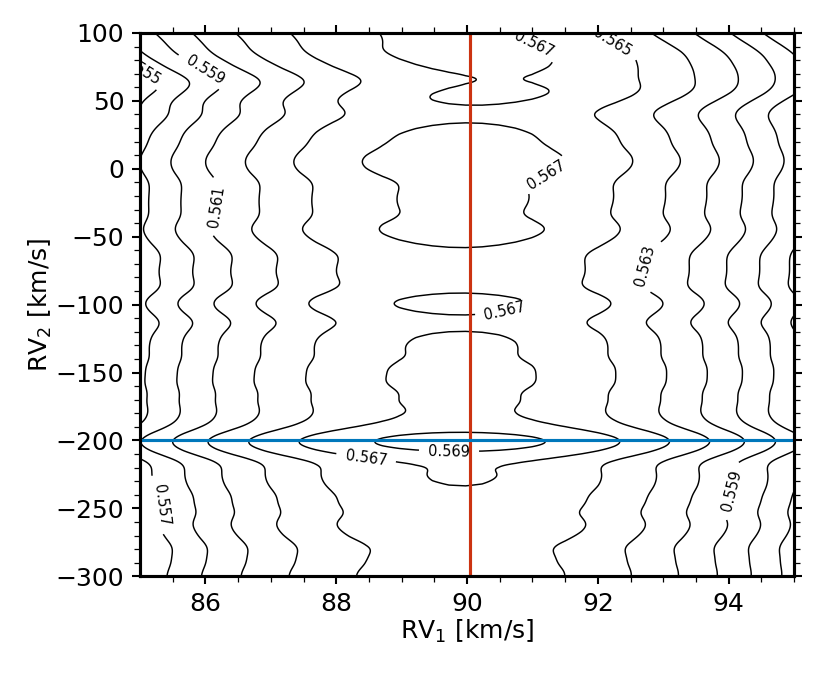}
\caption{Two dimensional cross-correlation function computed with \texttt{TODMOR} resulting from the injection-retrieval of a secondary spectra with a flux ratio of 2\%. The red vertical and blue horizontal lines indicate the RV position of the primary component and that of the injected signal, respectively. The numbers on the contours indicate the height of the CCF.}
\label{fig:2DCCF}
\end{figure}

\begin{figure} 
\centering
\includegraphics[width=0.5\textwidth]{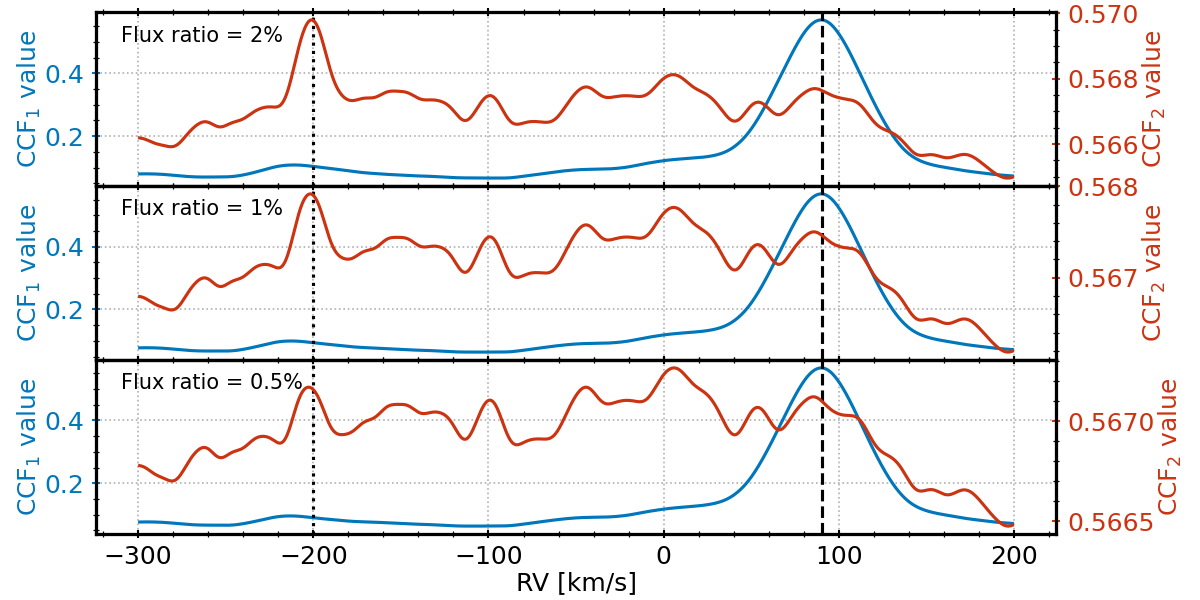}
\caption{Results from the injection-retrieval of secondary signals with flux ratios of 2\% (top), 1\% (middle) and 0.5\% (bottom) . The panels display one-dimensional sections of the two-dimensional CCF computed with \texttt{TODMOR}, corresponding to the vertical red and horizontal blue lines shown in Fig.~\ref{fig:2DCCF}. The  dashed and dotted vertical lines indicate the RV positions of the primary component and the injected signal, respectively.}
\label{fig:injection}
\end{figure}

We show in Fig.~\ref{fig:2DCCF} the two-dimensional cross correlation obtained with \texttt{TODMOR} for the injection of a spectrum with a flux ratio of 2\%, while in Fig.~\ref{fig:injection} we show the resulting one-dimensional sections of the  two-dimensional cross correlation,
computed for injected spectra with three different flux ratios. For the spectra with injected flux ratios of 2\% and 1\%, the CCF peak of the secondary component (red) is located at the radial velocity of the injected signal, and is easily detected. The signal produced by an injected spectrum with a flux ratio of 0.5\% is of the same height as some of the peaks of the surrounding noise, and therefore we consider that flux ratio as barely detectable.  


As a result of these tests, we constrain the flux ratio of the unseen companion of J1255 to be lower than 1\%. This upper limit rules out any 
main-sequence (MS) companion with a temperature larger than $\sim3700$\,K, which have typical masses of $\sim0.5$\,M$_{\odot}$ \citep{Cifuentes2020}.

\subsection{Searching for a faint wide component around J1255 with Palomar Adaptive Optics}
\label{sec:AO}

To search for a faint wide companion to J1255 we obtained high-resolution imaging 
of J1255 with the infrared high-resolution adaptive optics (AO) imaging at Palomar Observatory, with the PHARO instrument \citep{hayward2001} at the 200-inch PALOMAR telescope, behind the natural guide star AO system P3K \citep{dekany2013}, 
on 2021~Feb~23 UT in a standard 5-point quincunx dither pattern with steps of 5 arc-sec.  Each dither position was observed three times, offset in position from each other by 0.5 arc-sec for a total of 15 frames. 

The camera was in the narrow-angle mode with a full field of view of $\sim25$ arc-sec and a pixel scale of approximately $0.025$ arc-sec per pixel. Observations were made in the narrow-band Br-$\gamma$ filter $(\lambda_o = 2.1686; \Delta\lambda = 0.0326\mu$m) and in the narrow-band $H-cont$ filter $(\lambda_o = 1.668; \Delta\lambda = 0.018\mu$m), each with an integration time of 30 seconds per frame (150 seconds total on-source time).

\begin{figure}
\centering
\includegraphics[width=0.5\textwidth]
{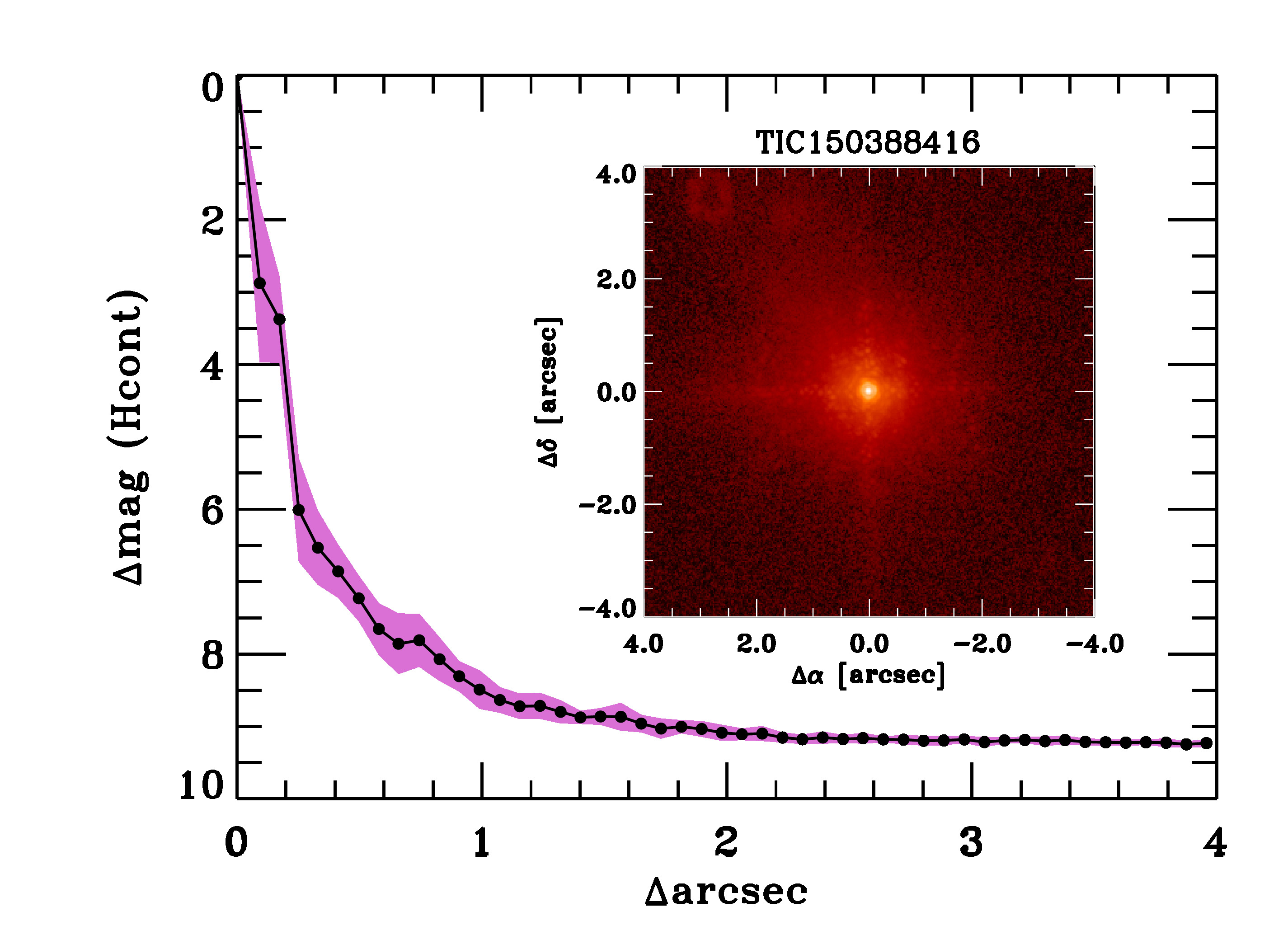}
\caption{Companion sensitivity for the Palomar adaptive optics imaging of the $H-cont$.  The black points represent the 5$\sigma$ limits and are separated in steps of 1 FWHM; the purple represents the azimuthal dispersion (1$\sigma$) of the contrast determinations (see text). The inset image is of the primary target, showing no additional companions to within $3$ arc-sec of the target in either filter.}
\label{fig:AO}
\end{figure}

The AO data were processed and analyzed with a custom set of IDL tools.  The science frames were flat-fielded and sky-subtracted.  The flat fields were generated from a median average of dark subtracted flats taken on-sky.  The flats were normalized such that the median value of the flats is unity.  The sky frames were generated from the median average of the 15 dithered science frames; each science image was then sky-subtracted and flat-fielded.  The reduced science frames were combined into a single combined image using an intra-pixel interpolation that conserves flux, shifts the individual dithered frames by the appropriate fractional pixels, and median-coadds the frames.  The final resolution of the combined dither was determined from the full-width half-maximum of the point spread function; $0.099$ arc-sec (60au) in Br-$\gamma$ and $0.082$ arc-sec (49 au) in $H-cont$ (Fig.~\ref{fig:AO} for  $H-cont$; similar results for Br-$\gamma$).

The sensitivities of the final combined AO image were determined by injecting simulated sources azimuthally around the primary target every $30^\circ $ at separations of integer multiples of the central source's FWHM \citep{furlan2017}. The brightness of each injected source was scaled until standard aperture photometry detected it with $5\sigma $ significance. The resulting brightness of the injected sources relative to the target set the contrast limits at that injection location. The final $5\sigma$ limit at each separation was determined from the average of all of the determined limits at that separation and the uncertainty on the limit was set by the rms dispersion of the azimuthal slices at a given radial distance (Fig.~\ref{fig:AO}).  The only other star detected is 12 arc-sec to the southwest; this star is TIC~150388417 (\textit{Gaia} DR3 1577114984684273664) and has a distance of 350~pc and is thus physically unrelated to J1255.   There is no evidence in either filter for an additional companion of J1255.

\bsp	
\label{lastpage}
\end{document}